\documentclass[preprintnumbers,prd,showpacs,floatfix,superscriptaddress,nofootinbib,twocolumn]{revtex4-2}

\usepackage{longtable}
\usepackage{bm}
\usepackage{relsize}
\usepackage{amsfonts}
\usepackage{amsmath}
\usepackage{amssymb,epsf}
\usepackage{latexsym}
\usepackage{graphicx,epsfig}
\usepackage{amssymb}
\usepackage{float}
\usepackage{subfigure}
\usepackage{epstopdf}
\usepackage[colorlinks=true,citecolor=blue,linkcolor=blue,urlcolor=black]{hyperref}
\usepackage{dcolumn}
\usepackage{psfrag}
\usepackage{wrapfig}
\usepackage{makeidx}
\usepackage{epsf}
\usepackage{color}
\usepackage{multirow}
\usepackage{mathtools}

\usepackage{tensor}
\usepackage{xcolor}

\begin{document}

\title{Gravitationally induced matter creation in scalar-tensor $f(R,T_{\mu\nu}T^{\mu\nu})$ gravity}

\author{Ricardo A. C. Cipriano}
\email{ricardocipriano98@gmail.com}
\affiliation{Instituto de Astrofísica e Ciências do Espaço, Faculdade de Ciências da Universidade de Lisboa, Edifício C8, Campo Grande, P-1749-016 Lisbon, Portugal}
\affiliation{Departamento de F\'{i}sica, Faculdade de Ci\^{e}ncias da Universidade de Lisboa, Edifício C8, Campo Grande, P-1749-016 Lisbon, Portugal}
\author{Tiberiu Harko}
\email{tiberiu.harko@aira.astro.ro}
\affiliation{Department of Physics, Babes-Bolyai University, Kogalniceanu Street,
	Cluj-Napoca 400084, Romania}
\affiliation{Department of Theoretical Physics, National Institute of Physics
and Nuclear Engineering (IFIN-HH), Bucharest, 077125 Romania}
\affiliation{Astronomical Observatory, 19 Ciresilor Street,
	Cluj-Napoca 400487, Romania}
\author{\\Francisco S. N. Lobo}
\email{fslobo@fc.ul.pt}
\affiliation{Instituto de Astrofísica e Ciências do Espaço, Faculdade de Ciências da Universidade de Lisboa, Edifício C8, Campo Grande, P-1749-016 Lisbon, Portugal}
\affiliation{Departamento de F\'{i}sica, Faculdade de Ci\^{e}ncias da Universidade de Lisboa, Edifício C8, Campo Grande, P-1749-016 Lisbon, Portugal}
\author{Miguel A. S. Pinto}
\email{mapinto@fc.ul.pt}
\affiliation{Instituto de Astrofísica e Ciências do Espaço, Faculdade de Ciências da Universidade de Lisboa, Edifício C8, Campo Grande, P-1749-016 Lisbon, Portugal}
\affiliation{Departamento de F\'{i}sica, Faculdade de Ci\^{e}ncias da Universidade de Lisboa, Edifício C8, Campo Grande, P-1749-016 Lisbon, Portugal}
\author{João Luís Rosa}
\email{joaoluis92@gmail.com}
\affiliation{Institute of Physics, University of Tartu, W. Ostwaldi 1, 50411 Tartu, Estonia}
\affiliation{University of Gda\'{n}sk, Jana Ba\.{z}y\'{n}skiego 8, 80-309 Gda\'{n}sk, Poland}

\date{\today}

\begin{abstract}
In this work, we analyze the possibility of gravitationally induced matter creation in the so-called Energy-Momentum-Squared gravity (EMSG), i.e. $f(R,T_{\mu\nu}T^{\mu\nu})$ gravity, in its dynamically equivalent scalar-tensor representation. Given the explicit nonminimal coupling between matter and geometry in this theory, the energy-momentum tensor is not generally covariantly conserved, which motivates the study of cosmological scenarios by resorting to the formalism of irreversible thermodynamics of open systems. We start by deriving the universe matter creation rates and subsequent thermodynamical properties, such as, the creation pressure, temperature evolution, and entropy evolution, in the framework of $f(R,T_{\mu\nu}T^{\mu\nu})$ gravity. These quantities are then analyzed for a Friedmann-Lemaître-Robertson-Walker (FLRW) background with a scale factor described by the de Sitter solution, under different assumptions for the mater distribution, namely a vacuum universe, a constant density universe, and a time-varying density universe. Finally, we explore cosmological solutions with varying Hubble parameters and provide a comparison with the standard cosmological model. Our results indicate that the cosmological evolution in the framework of EMSG are in close agreement with the observational cosmological data for low redshift.
\end{abstract}


\maketitle


\section{Introduction}\label{sec:intro}

For the past century, General Relativity (GR) \cite{Einstein:GR} has prevailed as the most successful description of the gravitational interaction. Indeed, Einstein's theory possesses astonishing predictions that have been observed throughout the years, such as the imaging of supermassive black hole candidates at the center of galaxies \cite{EventHorizonTelescope:2019dse,EventHorizonTelescope:2022wkp} and gravitational waves \cite{LIGOScientific:2016aoc,LIGOScientific:2017vwq,LIGOScientific:2017ync,KAGRA:2023pio}, which has fundamentally altered the way we perceive and test the Cosmos. Nevertheless, GR has also been shown to have some shortcomings, either directly from the observational data \cite{Planck:2015fie, Planck:2018vyg, WMAP:2003elm} or from a purely theoretical point of view \cite{Capozziello:2011et}. An example of the former is the discovery of the late-time cosmic accelerated expansion of the universe, found through observations of Type Ia supernovae \cite{SupernovaSearchTeam:1998fmf,SupernovaCosmologyProject:1998vns,Boomerang:2000efg,Hanany:2000qf, SupernovaCosmologyProject:2003dcn,Amanullah:2010vv}. Currently, the best attempt to explain such an evolutionary stage comes from the standard cosmological model, the $\Lambda$CDM model, which takes GR as its basis, with $\Lambda$ the cosmological constant introduced by Einstein \cite{Einb}. Albeit from a phenomenological point of view the $\Lambda$CDM model describes the universe with great accuracy, although from a theoretical point of view the theory entails some open questions such as the Cosmological Constant Problem (see \cite{Weinberg:1988cp,Carroll:2000fy} for a review on this issue) or the coincidence problem \cite{Bull:2015stt}, which leads us to question whether a more general theory of gravity could exist. Furthermore, GR encapsulates singularities, such as the primordial Big Bang singularity, which are still puzzling to the physics community, as it usually needs a quantum description of gravity to be fully understood \cite{Capozziello:2011et}.

In order to avoid the issues mentioned above, a common practice is to consider alternative theories of gravity, where a particular way to theoretically explain these problems without changing GR's core principles and theoretical framework is to consider Extended Theories of Gravity (ETG's) \cite{Capozziello:2011et}. Indeed, an alternative approach to the cosmological analysis is to consider GR as an approximation of a more complete theory of gravity \cite{Nojiri:2006gh, Nojiri:2007as}, which is consistent with the observational data. ETG's have been extensively studied in the literature during the past century, even almost immediately after the publication of GR, with the pioneering work developed by Weyl on attempting to gauge Einstein's theory \cite{Weyl:1917gp, Weyl:1919fi}, to $f(R)$ gravity, introduced in the 1970s by Buchdahl \cite{Buchdahl:1970ynr}. In fact, presently there is a plethora of ETG's \cite{Capozziello:2007ec,Clifton:2011jh}, each possessing a varying degree of success and specific characteristics. However, none fit the observational data better than the $\Lambda$CDM model.

In particular, a specific ETG, the so-called the Energy-Momentum-Squared gravity (EMSG), represented by a function that depends on the Ricci scalar $R$ and on the contraction between the energy-momentum tensor $T_{\mu\nu}$ with itself, $f(R,T_{\mu\nu}T^{\mu\nu})$ \cite{Katirci:2013okf,Roshan:2016mbt}, has recently received much attention.
Henceforth, we will define $\mathbf{T}^2 \equiv T_{\mu\nu}T^{\mu\nu}$ for notational simplicity and convenience.
In fact, EMSG has been of particular interest due to its astrophysical and cosmological implications in high curvature (high energy) regimes, and, as a consequence, it is possibly easier to find constraints directly from observations.
Interestingly, it was shown that the theory may avoid the primordial singularity with a bouncing cosmology, and it was subsequently shown in a non-quantum approach that these solutions could have a finite maximum energy density and a scale factor smaller than the minimum length scale \cite{Board:2017ign}. Although it has been argued that a universe which can regularly connect the early universe bounce to a viable de Sitter late-time universe should not generally exist \cite{Barbar:2019rfn}, it was pointed out that this issue is solved with the existence of a vacuum energy density in EMSG \cite{Nazari:2020gnu}.

Moreover, it is also important to note that even if EMSG avoids a primordial singularity it does not prevent the appearance of black hole solutions (but allows, for example, the existence of more massive neutron stars \cite{Nari:2018aqs,Akarsu:2018zxl}).
In this sense, charged black holes \cite{Chen:2019dip, Roshan:2016mbt} and neutron star solutions \cite{Akarsu:2018zxl} have been analyzed. Separately, the mass-radius relations of neutron stars, considering specific types of equations of state (EoS), were determined \cite{Tangphati:2022acb,Tangphati:2021wng}. The theoretical color-flavored quark stars were also studied in the context of EMSG \cite{Singh:2020bdv}, wormhole solutions were explored \cite{Moraes:2017dbs, Rosa:2023guo}, and the proprieties of black hole rings were analysed \cite{Chen:2019dip}. Furthermore, it was argued that the cosmic acceleration could be explained in a EMSG dust only universes \cite{Akarsu:2017ohj}, and an attempt to study a simple anisotropic model was explored \cite{Akarsu:2020vii}, as well as a dynamical-systems approach \cite{Bahamonde:2019urw}.

Regarding the theoretical aspects of EMSG, it is worth mentioning that it encodes a nonminimal coupling between the geometry of spacetime and matter. This explicit nonminimal coupling induces a non-vanishing covariant derivative of the energy-momentum tensor, that implies non-geodesic motion and consequently leads to the appearance of an extra force \cite{Bertolami:2007gv,Harko:2010mv,Haghani:2013oma,Harko:2014gwa,Harko:2014aja,Harko:2014sja,Harko:2018gxr,Harko:2010hw,Harko:2012hm,Bertolami:2008zh,Harko:2020ibn,Harko:2012ve}.
The cosmological implications of these geometry-matter nonminimal couplings have been extensively studied, where the important results reside in the capability of explaining the late-time cosmic acceleration as well as the evolution of the inflationary stage. As an important remark, EMSG may not be equivalent to $f(R)$ gravity when the trace of the energy-momentum tensor, $T=g^{\mu\nu}T_{\mu \nu}$, is zero. For example, in the case of the electromagnetic field, although $T=0$, in EMSG non-vanishing terms still exit \cite{Akarsu:2017ohj}, analogously to those that appear in loop-quantum gravity and braneworld solutions \cite{Ashtekar:2011ni, Brax:2003fv} respectively, whereas theories such as $f(R,T)$ gravity \cite{Harko:2011kv} do not have this propriety \cite{Akarsu:2017ohj}. 

From a theoretical point of view, theories of the type $f\left(R,L_m\right)$, $f(R,T)$ or $f\left(R,T_{\mu \nu}T^{\mu \nu}\right)$, can be considered as maximal extensions of the standard additive Hilbert-Einstein Lagrangian, $\mathcal{L}_{EH}=R/2\kappa^2+\mathcal{L}_m$, where $\mathcal{L}_m$ is the matter action. It is certainly interesting to go beyond the additive algebraic structure of the Lagrangian of standard GR, and look for gravitational theories having a more intricate algebraic structure in their Lagrangian, involving a non-standard interaction between matter and geometry. Theories with geometry-matter coupling have the intriguing property that energy-momentum tensor of the baryonic matter is not conserved. This phenomenon also appears when one considers quantum field theory in curved spacetimes, and as a quantum effect in the expanding Universe. Thus we are led to the interesting possibility that perhaps theories with geometry-matter coupling may somehow be related to the quantum aspects of gravity, and they may represent a (very small) step in the understanding of quantum gravity phenomenology. Moreover,  these classes of modified gravity theories give a good description of the cosmological and astrophysical observational data, and generally to the gravitational interactions at small and large scales.

In this context, this paper aims to study ESMG on two new fronts: an equivalent scalar-tensor approach, which has shown to possess interesting results in early-time cosmology and the subsequent inflation period in a different manner, as well as reducing the order of the field equations on the metric tensor; and an approach based on the work of Prigogine and coworkers \cite{Prigogine:1986,Prigogine:1988, Prigogine:1989zz} that attempts to theorize a way of creating matter in the universe via gravitational processes by using the formalism of irreversible thermodynamics of open systems. It has been shown that the irreversible thermodynamics of open systems possesses some interesting cosmological applications \cite{Lima:1992np,Calvao:1991wg,Graef:2013iia,Lima:2014qpa,Harko:2014pqa,Lima:2014hda, Harko:2015pma,Harko:2021bdi,Pinto:2022tlu,Pinto:2023tof,Pinto:2023phl, Pinto:2023}. In fact, constructing an ETG with dissipative processes leads to a gravitational entropy enticed with the creation of matter that can occur from initial empty conditions which can be falsified by fundamental particle physics \cite{Harko:2014pqa}. As a further note, a similar approach was recently taken in \cite{Pinto:2022tlu} and \cite{Pinto:2023} for the $f(R,T)$ gravity, in which the description of the matter creation processes lead to a generalization of the $\Lambda$CDM model.

This paper is organized as follows. In Sec. \ref{sec:theory} we introduce the action and equations of motion for EMSG and derive the corresponding scalar-tensor representation for the field equations. In Sec. \ref{sec:thermodynamics}, we present the formalism of irreversible thermodynamics of open systems and obtain a set of several quantities of interest in this formalism, such as, the particle creation rate and the creation pressure. In Sec. \ref{sec:cosmology}, we introduce a FLRW background and obtain the corresponding cosmological equations for the theory, as well as the relevant thermodynamic variables, in this particular background. In Sec. \ref{sec:models}, we analyze cosmological models under this framework, in particular the de Sitter solution and solutions with a varying Hubble parameter. Finally, in Sec. \ref{sec:concl}, we summarize and discuss our results.

\section{Theory and framework}\label{sec:theory}

\subsection{Geometrical representation}

We assume that the equations of motion, as well as the thermodynamic properties of the baryonic matter can be described by means of a matter Lagrangian density $\mathcal{L}_m$.  Moreover, we introduce the matter energy-momentum tensor, defined according to the prescription
\begin{equation}
T_{\mu \nu} \equiv -\frac{2}{\sqrt{-g}} \frac{\delta\left(\sqrt{-g} \mathcal{L}_m\right)}{\delta g^{\mu \nu}},
\end{equation}
where  $g$ is the determinant of the metric, and $\delta\left(\sqrt{-g} \mathcal{L}_m\right)/\delta g^{\mu\nu}$ is the functional derivative of $\sqrt{-g} \mathcal{L}_m$ with respect to $g^{\mu\nu}$.

As mentioned above, one of the many possible extensions of the $f(R)$ gravity theory is to consider a more general function, $f(R,\mathbf{T}^2)$, where $R$ is the Ricci scalar $R=g^{\mu \nu} R_{\mu \nu}$, $R_{\mu \nu}$ is the Ricci tensor, and $\mathbf{T}^2$ is the contraction of the energy-momentum tensor $T_{\mu\nu}$, $\mathbf{T}^2 \equiv T_{\mu\nu}T^{\mu\nu}$.
Such an extension accounts for the matter content of the Universe and its possible nonminimal coupling to the geometry, offering a more general description of the gravitational interaction. The action $S$, which describes the $f(R, \mathbf{T}^2)$ gravity theory is given by \cite{Katirci:2013okf,Roshan:2016mbt}
\begin{equation}
S=\frac{1}{2 \kappa^2} \int_{\Omega} \sqrt{-g} f(R, \mathbf{T}^2) d^4 x+\int_{\Omega} \sqrt{-g} \mathcal{L}_m d^4 x \,,
\label{eq1_1}
\end{equation}
where $\kappa^2=\kappa^2 G / c^4$,  $G$ is the gravitational constant, $c$ is the speed of light in vacuum, $\Omega$ is the 4-dimensional Lorentzian manifold that represents spacetime, and in which one defines a set of coordinates $\{x^\mu\}$, In addition, $f(R,\mathbf{T}^2)$ is an arbitrary well-behaved function of the Ricci scalar $R$, the scalar $\mathbf{T}^2$.

Furthermore, to simplify the notation, henceforth, we consider a system of geometrized units for which $G=c=1$, which implies that $\kappa^2=8\pi$. Under the metric formalism, in the geometrical representation, the only field that mediates the gravitational interaction is the metric tensor $g_{\mu\nu}$. Thus, the field equations of $f(R, \mathbf{T}^2)$ gravity are obtained by taking the variation of Eq. \eqref{eq1_1} with respect to $g_{\mu \nu}$, yielding
\begin{equation}
f_R R_{\mu \nu}-\frac{1}{2} g_{\mu \nu} f-\left(\nabla_\mu \nabla_\nu-g_{\mu \nu} \square\right) f_R=\kappa^2 T_{\mu \nu}-f_{\mathbf{T}^2} \Theta_{\mu \nu},
\label{eq3_1}
\end{equation}
where we have defined $f_R \equiv \partial f / \partial R$ and $f_{\mathbf{T}^2} \equiv \partial f / \partial \mathbf{T}^2$ and introduced $\Theta_{\mu \nu}$ as an auxiliary tensor, defined as
\begin{equation}
\Theta{\mu \nu} \equiv \frac{\delta \left(T^{\alpha \beta} T{\alpha \beta}\right)}{\delta g^{\mu \nu}},
\label{eq4_1}
\end{equation}
whose explicit form will be settled once either the energy-momentum tensor $T_{\mu \nu}$ or, equivalently, the matter Lagrangian density $\mathcal{L}_m$ are specified. The conservation equation can be obtained by taking the covariant derivative of Eq. \eqref{eq3_1}, resulting in
\begin{equation}
\kappa^2 \nabla_\mu T^{\mu \nu}=\nabla_\mu\left(f_{\mathbf{T}^2} \Theta^{\mu \nu}\right)+f_R \nabla_\mu R^{\mu \nu}-\frac{1}{2} g^{\mu \nu} \nabla_\mu f.
\label{eq5_1}
\end{equation}
It is important to emphasize that the conservation of the $T_{\mu\nu}$, i.e., $\nabla_\mu T^{\mu \nu}=0$ is no longer a requirement of the theory, unlike in GR or $f(R)$ gravity for instance. Nevertheless, such a condition is commonly considered as an extra assumption in cosmological models \cite{Goncalves:2021ybs,Goncalves:2023klv,Goncalves:2022ggq,Goncalves:2021vci}.

\subsection{Scalar tensor representation}
Similarly to what happens in other modified theories of gravity featuring extra scalar degrees of freedom in comparison with GR, $f(R, \mathbf{T}^2)$ gravity admits a transformation into a dynamically equivalent scalar-tensor theory, in which the arbitrary dependency of the action in $R$ and $\mathbf{T}^2$ are exchanged by an arbitrary dependency on two scalar fields. To perform such a transformation, we introduce two auxiliary fields $\alpha$ and $\beta$ in the geometrical part of the action as
\begin{eqnarray}
S=\frac{1}{2 \kappa^2} \int_{\Omega} \sqrt{-g}\Big[f(\alpha, \beta)+f_\alpha(R-\alpha)
    \nonumber \\
+f_\beta(\mathbf{T}^2-\beta)\Big] d^4 x,
\label{eq6}
\end{eqnarray}
where we have defined $f_\alpha \equiv \partial f / \partial \alpha$ and $f_\beta \equiv \partial f / \partial \beta$. The action in Eq. \eqref{eq6} now depends on three fundamental fields, namely, the metric $g_{\mu \nu}$ and the auxiliary fields $\alpha$ and $\beta$. Then, taking the variation of Eq. \eqref{eq6} with respect to $\alpha$ and $\beta$ yields the system of equations
\begin{subequations}
\begin{align}
f_{\alpha \alpha}(R-\alpha)+f_{\beta \alpha}(\mathbf{T}^2-\beta)&=0, \label{eq7}\\
f_{\alpha \beta}(R-\alpha)+f_{\beta \beta}(\mathbf{T}^2-\beta)&=0. \label{eq8}
\end{align}
\end{subequations}

The system of Eqs. \eqref{eq7} and \eqref{eq8} can be rewritten in terms of a matrix equation of the form $\mathcal{M} \mathrm{x}=0$ as
\begin{equation}
\mathcal{M} \mathrm{x}=\left(\begin{array}{ll}
f_{\alpha \alpha} & f_{\beta \alpha} \\
f_{\alpha \beta} & f_{\beta \beta}
\end{array}\right)\left(\begin{array}{l}
R-\alpha \\
\mathbf{T}^2-\beta
\end{array}\right)=0.
\label{eq9}
\end{equation}
The solution of a matrix system of the form of Eq. \eqref{eq9} is unique if and only if the determinant of the matrix $\mathcal{M}$ does not vanish, i.e., $\operatorname{det} \mathcal{M} \neq 0$. For any function $f(\alpha, \beta)$ satisfying the Schwarz theorem, i.e. $f_{\alpha \beta}=f_{\beta \alpha}$, this condition implies
\begin{equation}
f_{\alpha \alpha} f_{\beta \beta} \neq f^2_{\alpha \beta}.
\label{eq10}
\end{equation}
If Eq. \eqref{eq10} is satisfied, then the solution of Eq. \eqref{eq9} is unique and it is given by $\alpha=R$ and $\beta=\mathbf{T}^2$. Replacing this solution back into Eq. \eqref{eq6}, one recovers the geometrical part of Eq. \eqref{eq1_1}, thus proving that the two actions are equivalent. However, if Eq. \eqref{eq10} is not satisfied, then the equivalence of both representations is not guaranteed.

One can now introduce the definitions of the scalar fields $\phi$ and $\psi$,
\begin{equation}
\phi \equiv \frac{\partial f}{\partial \alpha}, \qquad \psi \equiv \frac{\partial f}{\partial \beta},
\label{eq11}
\end{equation}
respectively, as well as the interaction potential $V$,
\begin{equation}
V(\phi, \psi)=-f(\alpha, \beta)+\phi \alpha+\psi \beta,
\label{eq12}
\end{equation}
which, upon a replacement into Eq. \eqref{eq6}, yields the action of the scalar-tensor representation of $f(R, \mathbf{T}^2)$ as
\begin{eqnarray}
S=\frac{1}{2 \kappa^2} \int_{\Omega} \sqrt{-g}[\phi R+\psi \mathbf{T}^2-V(\phi, \psi)] d^4 x
    \nonumber \\
+\int_{\Omega} \sqrt{-g} \mathcal{L}_m d^4 x.
\label{eq13}
\end{eqnarray}
Equation \eqref{eq13} depends now on three independent quantities, namely the metric $g_{\mu \nu}$ and the scalar fields $\phi$ and $\psi$.

A variation with respect to the metric $g_{\mu \nu}$ yields the field equations in the scalar-tensor representation as
\begin{eqnarray}
\phi G_{\mu \nu} +\frac{1}{2} g_{\mu \nu} V-\left(\nabla_\nu \nabla_\mu-g_{\mu \nu} \square\right) \phi
	\nonumber \\
=\kappa^2 T_{\mu \nu}-\psi\left(\Theta_{\mu \nu}-\frac{1}{2} g_{\mu \nu} \mathbf{T}^2\right)\,.
\label{eq14}
\end{eqnarray}
We have introduced the Einstein tensor $G_{\mu \nu}=$ $R_{\mu \nu}-(1/2) R g_{\mu \nu}$, whereas a variation with respect to $\phi$ and $\psi$ yields the equations of motion for these fields respectively as
\begin{subequations}
\begin{align}
V_\phi &= R, \label{eq15}\\
V_\psi &= \mathbf{T}^2,\label{eq16}
\end{align}
\end{subequations}
where we have defined $V_\phi \equiv \partial V / \partial \phi$ and $V_\psi \equiv \partial V / \partial \psi$.

Finally, taking a covariant derivative of Eq. \eqref{eq14}, one obtains the conservation equation in the scalar-tensor representation, given by
\begin{equation}
\kappa^2 \nabla_\mu T^{\mu \nu}=\nabla_\mu\left(\psi \Theta^{\mu \nu}\right)-\frac{1}{2} g^{\mu \nu}\left[R \nabla_\mu \phi+\nabla_\mu(\psi \mathbf{T}^2-V)\right].
\label{eq17}
\end{equation}
Note that inserting the definitions given in Eqs. \eqref{eq11} and \eqref{eq12} into Eqs. \eqref{eq14} and  \eqref{eq17} one recovers Eqs. \eqref{eq3_1} and \eqref{eq5_1}, which emphasizes the equivalence between the two representations.

\section{Thermodynamics of open systems}\label{sec:thermodynamics}

The study of irreversible matter creation in the realm of cosmology traces its roots back to the pioneering work of Prigogine and collaborators \cite{Prigogine:1986,Prigogine:1988,Prigogine:1989zz}. In the formalism established in these papers, the universe is viewed as an open system, and particle creation is described through the integration of a matter creation term in the energy-momentum tensor which leads to a reinterpretation of the conservation laws (we refer the reader to Ref. \cite{Pinto:2023phl} for a review). In this Section, we delve into the thermodynamics of open systems and apply this formalism to an homogeneous and isotropic universe.

\subsection{Matter creation in cosmology}

There are several known physical mechanisms, mostly appearing in quantum field theory in curved spacetimes, which allow the production of particles in various gravitational backgrounds \cite{Parker:1968mv,Parker:1969au,Parker:1971pt,Parker:1972kp}. The best known of these processes is particle production in curved spacetimes, which can be briefly summarized as follows (see \cite{Haro:2010mx,Haro:2018zdb}, and references therein). The Lagrangian of a scalar field is given by
\begin{equation}
{\cal L}=\frac{1}{2}\left( \nabla_\mu \nabla^\mu - m^2 \phi^2 -\xi R \phi ^2 \right),
\end{equation}
for constant $m$ and $\xi$, giving for the evolution of the scalar field the generalized Klein-Gordon equation
\begin{equation}
\left(-\nabla _\mu\nabla ^\mu+m^2+\xi R\right)\phi=0.
\end{equation}

From the Klein-Gordon equation one can obtain the particle number density as produced by the expansion of the universe as given, in the adiabatic approximation, as $n=mH^2/512 \pi$, with the corresponding energy density given by $\rho =m^2H^2/96\pi$ \cite{Haro:2010mx}.

Particle creation processes may also appear as a result of the vacuum instability in background gravitational and gauge fields, and they may be due to the conformal trace anomaly
$\Gamma =(\pi /2)\left<T_\mu^\mu\right>$, where by $\left<T_\mu^\mu\right>$ we have denoted the anomalous trace of the energy-momentum tensor $T^{\mu \nu}$ \cite{Chernodub:2023pwf}.  This formula can reproduce the radiation generated by static gravitational fields, and describe Schwinger pair creation in massless (scalar and spinor) quantum
electrodynamics, as well as  the photon and neutrino pair production \cite{Chernodub:2023pwf}. Hence, there are a large number of physical processes that may generate particles via various, mostly quantum mechanisms, and the approach considered in the present work may give some insights, on a classical level, of the purely quantum processes that may play also a dominant role in cosmology.

Particles creation processes may also be present in Weyl geometric approaches to gravitational theories, due
to the existence of quadratic curvature terms in the action, and of the direct interaction the perfect
fluid particles. Hence, in these models particles may be created directly from the vacuum \cite{Berezin1}.  Assuming that the particle creation rates are conformally invariant, the continuity equation takes the form $\nabla _\mu \left(nu^\mu\right)=\Phi$, where $u^\mu$ is the fluid velocity, and $\Phi$ is the creation rate. As for the matter creation rate, it is given by the quadratic expression \cite{Berezin1}
\begin{equation}
\Phi=\alpha_1'R_{\mu\nu\lambda\sigma}R^{\mu\nu\lambda\sigma}
+\alpha_2'R_{\mu\nu}R^{\mu\nu}
+\alpha_3'R^2+ \alpha_4'F_{\mu\nu}F^{\mu\nu},
\end{equation}
where $\alpha _i'$, $i=1,2,3,4$ are constants, and $F_{\mu \nu}$ is the strength of the Weyl vector. In the case of Weyl geometry, the particle creation law reduces to \cite{Berezin1}
\begin{equation}
\nabla _{\mu}\left(nu^\mu\right)=\eta C^2.
\end{equation}
where $\eta $ is a constant, while $C^2$ is the square of the Weyl tensor, $C^2=C_{\alpha \beta \gamma \delta}C^{\alpha \beta \gamma \delta}$. An interesting problem is related to the question that if the particles are created  from a vacuum state, the vacuum may persist or not. In the quadratic Weyl model of particle creation it was shown that the vacuum persist with respect to the production of matter with positive pressure. On the other hand, the vacuum becomes unstable once dust particles are produced.

A cosmological model in which the classical Friedman equations, describing the variation of $H$, the dark energy component,
and the matter densities are coupled to quantised field equations for massive scalar modes $M\gg H$ was considered in \cite{Xue}. As a result of particle production  a massive pair plasma state is formed on a macroscopic time scale, leading to the interaction of matter and dark energy densities. There is an alternating interaction between  dark energy and massive pairs that slows inflation to its end, and initiates the reheating process, by producing stable and unstable particle pairs.

\subsection{Thermodynamic interpretation of matter creation}

In a universe where matter can be created irreversibly, it is necessary and convenient to address it as an open system \cite{Prigogine:1989zz}, where the number of particles $N$ in a specific volume $V$ is not always fixed. When an adiabatic transformation occurs (a change in which no heat is exchanged with the environment) the thermodynamical energy conservation for open systems is given by
\begin{equation}
\label{1st law}
d(\rho V)+p d V=\frac{h}{n} d(n V),
\end{equation}
where $n=N / V$ is the particle number density and $h=\rho+p$ is the enthalpy per unit volume, with $p$ the pressure and $\rho$ the energy density. Following this, one can now introduce the second law of thermodynamics in the case of an open system which takes the form of
\begin{equation}
d \mathcal{S}=d_e \mathcal{S}+d_i \mathcal{S} \geq 0,
\label{dif_entro}
\end{equation}
where $d_e \mathcal{S}$ and $d_i \mathcal{S}$ are the entropy flow and entropy creation, respectively. A consequence of Eq. \eqref{dif_entro} is that the total entropy of the system and its surroundings either increases or stays the same over time.
In addition, it is possible to obtain an expression for the entropy flow and for the entropy creation \cite{Pinto:2022tlu}, which assume the following form, respectively
\begin{subequations}
\begin{align}
d_e \mathcal{S} &= \frac{d Q}{\mathcal{T}}, \label{eq9a}\\
d_i \mathcal{S} &= \frac{s}{n} d(n V) \label{eq9b}.
\end{align}
\end{subequations}
Due to the fact that, in an homogeneous universe, all physical quantities are independent of spacial coordinates and depend solely on time, such a universe cannot receive energy in the form of heat, i.e., $dQ=0$, and the 2nd law of thermodynamics becomes
\begin{equation}
d \mathcal{S}=\frac{s}{n} d(n V) \geq 0.
\label{dS_final}
\end{equation}
Hence, in a homogeneous universe, the change in entropy is sourced entirely by particle creation adiabatic processes. Additionally, Eq. \eqref{dS_final} suggests that cosmological matter can be created from the geometry of spacetime, but the inverse is not thermodynamically possible.

Let us now apply the non-equilibrium thermodynamics of open systems to cosmology. In this work, we consider a flat homogeneous and isotropic universe, that has a volume $V$ containing $N$ particles, an energy density $\rho$ and thermodynamic pressure $p$, which is well-described by the spatially flat Friedmann-Lemaître-Robertson-Walker (FLRW) metric with scale factor $a(t)$. We assume that the thermodynamic quantities describing baryonic matter satisfy the weak, the strong, and the dominant energy conditions, so that
\begin{equation}
\rho \geq 0, p\geq 0, \rho+p \geq 0, \rho+3p\geq 0, \rho \geq \left|p\right|.
\end{equation}

Under these assumptions, it is possible to express the comoving volume in terms of the scale factor, $V=a^3(t)$. Thus, Eq. \eqref{1st law} may be written as
\begin{equation}
\frac{d}{d t}\left(\rho a^3\right)+p \frac{d}{d t} a^3=\frac{\rho+p}{n} \frac{d}{d t}\left(n a^3\right),
\label{uau1}
\end{equation}
which can, in turn, be further simplified by introducing the Hubble function, $H=\dot{a} / a$, yielding
\begin{equation}
\dot{\rho}+3 H(\rho+p)=\frac{\rho+p}{n}(\dot{n}+3 H n).
\label{uau2}
\end{equation}
Eq.~\eqref{uau2} implies two major consequences. First, it plays the role of the effective conservation equation for open, homogeneous, and isotropic universes; and it implies that the ``heat'' received is caused solely by the variation in the particle number density, $n$. To further continue with this analysis it is useful to introduce the number current, which is defined as
\begin{equation}
\label{number_current}
N^\mu\equiv n u^\mu,
\end{equation}
where $u^\mu$ is the 4-velocity of the fluid which satisfies the normalization condition $u^\mu u_\mu=-1$. For a comoving observer, the 4-velocity becomes $u^\mu = (1,0,0,0)$, and, consequently, the number current is given by $N^\mu = (n,0,0,0)$. Accordingly, in such a frame, the covariant divergence of the number current is given by
\begin{equation}
\label{cov_nd}
\nabla_\mu N^\mu =  \dot{n}+3Hn \equiv  n \Gamma,
\end{equation}
where $\Gamma$ denotes the matter creation rate. The matter creation rate is defined according to the general relation
\begin{equation}
\Gamma =\frac{1}{N}\frac{dN}{dt}=\frac{1}{nV}\frac{d}{dt}(nV),
\end{equation}
and, from a physical point of view, it gives the normalized variation of the particle number in a given volume.

By substituting Eq. \eqref{cov_nd} into Eq. \eqref{uau2} we obtain the energy conservation equation in an alternative form
\begin{equation}
\dot{\rho}+3 H(\rho+p)=(\rho+p) \Gamma,
\label{uau4}
\end{equation}
or
\begin{equation}\label{31}
\Gamma =3H+\frac{\dot{\rho}}{\rho +p}.
\end{equation}

When the particle creation rate $\Gamma =0$, that, in the absence of particle creation, we recover the standard conservation law of GR. It is interesting to note that when $\rho =p=0$, the energy conservation equation is identically satisfied. Another interesting limiting case corresponds to a system in the presence of matter creation, but in which the particle number density is a constant, $\dot{n}=0$, or, equivalently, $\dot{\rho}=0$. In this case, as one can see from Eqs.~(\ref{cov_nd}) and (\ref{31}), the particle creation rate $\Gamma =3H$. This case corresponds to an expanding Universe in which  matter creation exactly compensates the variation of particle number density due to the cosmic expansion.

Moreover, it is possible to write Eq. \eqref{uau1} in terms of an additional pressure, namely the creation pressure (denoted by $p_{\mathrm{c}}$), that must be included, alongside $p$, in the generalized thermodynamic pressure of the open systems $\tilde{p}$, such that $p \Rightarrow \widetilde{p} \equiv p+p_c$. This consideration allows us to rewrite the 1st law of thermodynamics in the context of matter creation in an open system as
\begin{equation}
\frac{d}{d t}\left(\rho a^3\right)+\left(p+p_c\right) \frac{d}{d t} a^3=0,
\end{equation}
making it possible to write an expression for the creation pressure that depends solely on the creation rate
\begin{equation}
p_c=-\frac{\rho+p}{3 H} \Gamma.
\label{pc}
\end{equation}
Finally, Eq. \eqref{uau4} can be rewritten to obtain an explicit form for the creation rate
\begin{equation}
\Gamma=\frac{1}{\rho+p}[\dot{\rho}+3 H(\rho+p)].
\label{gama}
\end{equation}

\subsection{Temperature and entropy evolution}

Another advantage of using the formalism of irreversible thermodynamics of open systems in theories with a non-conservation of the matter energy-momentum tensor is the possibility of having a universe starting from empty conditions and gradually generating entropy with time. As such, we explore such an evolution, and also a temperature evolution.

For this analysis we define the entropy current $\mathcal{S}^\mu$ as
\begin{equation}
\label{entropy_current}
\mathcal{S}^\mu=s u^\mu,
\end{equation}
which in the comoving frame reduces to $\mathcal{S}^\mu=(s,0,0,0)$. The covariant divergence of the entropy current assumes the following form \cite{Lima:2014hda}
\begin{equation}
\nabla_\mu \mathcal{S}^\mu =n \dot{\sigma}+\sigma n \Gamma,
\end{equation}
where $\sigma=\mathcal{S}/N$ is the entropy per particle.
The expression for the entropy evolution can be obtained by extending the reasoning surrounding Eq. \eqref{dif_entro}, and comparing it to Eq. \eqref{uau4} to obtain the following
\begin{equation}
\dot{\mathcal{S}}=\mathcal{S} \Gamma \geq 0.
\label{nhe1}
\end{equation}

Under this differential form, Eq. \eqref{nhe1} has the general solution
\begin{equation}\label{33}
\mathcal{S}(t)=\mathcal{S}_0 \exp \left[\int_0^t \Gamma\left(t^{\prime}\right) d t^{\prime}\right],
\end{equation}
where $\mathcal{S}_0=\mathcal{S}(0)$ is the constant initial entropy.

 As one can see from the above equation, in the thermodynamic formalism considered in the present work, which assumes a constant entropy per particle,  the generation of gravitational energy from matter annihilation, corresponding to $\Gamma <0$,  is forbidden by the second law of thermodynamics. More exactly, if in  Eq. (\ref{33}), which expresses the entropy in terms of the matter creation rate only, one assumes  $\Gamma<0$, the entropy would be a decreasing function of time, a result which would contradict the second law of thermodynamics. Thus, in the present approach, we assume that the second law of thermodynamic is valid for all matter forms, and restrict our investigations to a positive particle creation rates.

It is also possible to obtain an expression for the temperature evolution since we know that a thermodynamic system is fundamentally described by the particle number density $n$ and the temperature $\mathcal{T}$ as
\begin{equation}
\rho=\rho(n, \mathcal{T}), \qquad p=p(n, \mathcal{T}),
\end{equation}
which allows us to describe the differential of the energy density. Taking into account Eq. \eqref{uau4}, one thus obtains
\begin{equation}
\left(\frac{\partial \rho}{\partial n}\right)_{\mathcal{T}} \dot{n}+\left(\frac{\partial \rho}{\partial \mathcal{T}}\right)_n \dot{\mathcal{T}}+3(\rho+p) H=(\rho+p) \Gamma.
\label{cons_new}
\end{equation}

It has been pointed out in \cite{Saridakis} that if dark energy has an arbitrary, varying equation-of-state parameter $w(a)$,  all the cosmological quantities are well defined and regular for every $w(a)$, including at the $-1$-crossing. However, interestingly enough, the temperature is negative in the phantom regime $w(a)<-1$,  and positive in the quintessence phase, $w(a)>-1$. On the other hand, the matter density and the entropy are always positive. The temperature negativity can be related to some quantum effects, which may also appear in black holes \cite{Norte}. Negative temperatures have been shown to exist in a number of spin systems, and recently experiments have indicated
the existence of negative temperatures in atomic systems in their motional degrees of freedom \cite{Norte}. The regular behavior of all quantities at the $-1$-crossing suggests that such a crossing does not correspond to a phase transition, but to a smooth cross-over \cite{Saridakis}.

One can now obtain the explicit thermodynamical relation for $\left(\partial \rho / \partial n\right)_{\mathcal{T}}$ and, considering Eq. \eqref{uau4}, leads to the following expression for the temperature evolution
\begin{equation}
\frac{\dot{\mathcal{T}}}{\mathcal{T}}=c_s^2(\Gamma-3 H),
\label{mehhh}
\end{equation}
where we naturally defined the speed of sound $c_s=\sqrt{(\partial p / \partial \rho)_n}$. Equation \eqref{mehhh} has then the general solution of
\begin{equation}
\mathcal{T}(t)=\mathcal{T}_0 \exp \left\{c_s^2 \int_0^{t^{\prime}}\left[ \Gamma\left(t^{\prime}\right)-3 H\right] d t^{\prime}\right\},
\end{equation}
where $\mathcal{T}_0=\mathcal{T}(0)$ is the constant initial temperature.

\section{Cosmological equations}\label{sec:cosmology}

With all the preparation on the previous sections, we now explore cosmological models for the scalar-tensor representation of $f(R,\mathbf{T}^2)$ gravity that allow for matter creation. We assume the geometry of the universe to be well described by a homogeneous and isotropic FLRW model, for which the line-element can be written in the usual spherical coordinates $(t, r, \theta, \varphi)$ as
\begin{equation}
d s^2=-d t^2+a(t)^2\left[\frac{d r^2}{1-k r^2}+r^2\left(d \theta^2+\sin ^2 \theta d \varphi^2\right)\right],
\label{eq18}
\end{equation}
where $a(t)$ is the scale factor of the universe and $k$ is the spacial curvature parameter, where $k=$ $\{-1,0,1\}$ correspond to hyperbolic, flat, and spherical models, respectively. To preserve the isotropy of the spacetime, we assume that the two scalar fields $\phi \equiv \phi(t)$ and $\psi \equiv \psi(t)$ are functions solely of the time coordinate $t$.

We also assume that the matter content of the universe can be well described by an isotropic perfect-fluid, i.e., the matter energy-momentum tensor $T_{\mu \nu}$ can be written in the form
\begin{equation}
T_{\mu \nu}=(\rho+p) u_\mu u_\nu+p g_{\mu \nu},
\label{eq19}
\end{equation}
where $\rho \equiv \rho(t)$ is the energy density of the universe and $p \equiv p(t)$ is the isotropic pressure, both quantities assumed to depend solely on the time $t$ to preserve the isotropy of the spacetime, and $u^\mu$ is the 4-vector velocity of the fluid which satisfies the normalization condition $u_\mu u^\mu=-1$.

The matter Lagrangian density associated with the energy-momentum tensor $T_{\mu \nu}$ given in Eq. \eqref{eq19} is assumed in this work to be $\mathcal{L}_m=p$ \cite{Bertolami:2008ab}, and consequently the auxiliary tensor $\Theta_{\mu \nu}$ is given by
\begin{equation}
\Theta_{\mu \nu}=-\left(\rho^2+4 \rho p+3 p^2\right) u_\mu u_\nu.
\label{eq20}
\end{equation}
Under the assumptions outlined above, the field equations in Eq. \eqref{eq17} feature only two independent components, which correspond to the modified cosmological equations, i.e., the modified Friedmann and Raychaudhuri equations, which respectively take the forms
\begin{eqnarray}
3\left(H^2+\frac{k}{a^2}\right) &=& \frac{1}{\phi}\Big[\kappa^2 \rho+\frac{1}{2} V
	\nonumber \\	
	&& \hspace{-1cm} +\frac{1}{2} \psi\left(\rho^2+8 \rho p+3 p^2\right)-3 H \dot{\phi}\Big] \,,
\label{eq21}\\
2\left(\dot{H}-\frac{k}{a^2}\right) &=& \frac{1}{\phi}\Big[-\kappa^2(\rho+p)
	\nonumber \\
	&& \hspace{-1cm}  -\psi\left(\rho^2+4 \rho p+3 p^2\right)+H \dot{\phi}-\ddot{\phi}\Big] \,,
	\label{eq22}
\end{eqnarray}
where $H$ is again the Hubble parameter $H \equiv$ $\dot{a} / a$, and as usual the overdot $(\text{ }\dot{}\text{ } \equiv d / d t)$ denotes derivatives with respect to time. The equations of motion for the scalar fields $\phi$ and $\psi$ in Eqs. \eqref{eq15} and \eqref{eq16} become, respectively
\begin{subequations}
\begin{align}
V_\phi &= R =6\left(\dot{H}+2 H^2+\frac{k}{a^2}\right), \label{eq23}\\
V_\psi &= \mathbf{T}^2 = \rho^2+3 p^2. \label{eq24}
\end{align}
\end{subequations}

Finally, taking into account the expressions for the potential, the conservation equation (\ref{eq17}) takes the form
\begin{align}
\kappa^2[\dot{\rho}&+3 H(\rho+p)]+(\rho+p)(\rho+3 p)(\dot{\psi}+3 H \psi) \nonumber \\
&+\psi[\dot{\rho}(\rho+4 p)+\dot{p}(4 \rho+3 p)]=0.
\label{eq25}
\end{align}

When considering GR or any other modified theory of gravity in which the matter energy-momentum tensor $T_{\mu\nu}$ is conserved, it is common to assume that any non-equilibrium terms are incorporated in $T_{\mu\nu}$ \cite{Lima:2014hda}. However, since in $f(R,\mathbf{T}^2)$ gravity, the energy-momentum tensor is not necessarily conserved, such non-equilibrium terms, which are exclusively matter creation terms in our considerations, can be associated with this non-conservation. Therefore, by substituting Eq. \eqref{eq25} in Eq. \eqref{gama}, the matter creation rate takes the form
\begin{equation}
\Gamma =-\dfrac{1}{\kappa ^{2}}\bigg[(\dot{\psi}+3H\psi )(\rho +3p)+\psi
\frac{\dot{\rho}(\rho +4p)+\dot{p}(4\rho +3p)}{(\rho +p)}\bigg]
\label{creation_rate}
\end{equation}
whereas the creation pressure in Eq. \eqref{pc} becomes
\begin{equation}
p_{c}=\dfrac{\rho +p}{3H\kappa ^{2}}\bigg[(\dot{\psi}+3H\psi )(\rho
+3p)+\psi \frac{\dot{\rho}(\rho +4p)+\dot{p}(4\rho +3p)}{(\rho
+p)}\bigg].
\end{equation}%
These are the key parameters necessary to describe matter creation in $f(R,\mathbf{T}^2)$ gravity.
It is interesting to note that these quantities do  not depend on the spatial curvature $k$.

 The general behavior of the matter density in the present model follows from Eq.~(\ref{eq21}), and it is obtained as a solution of the algebraic equation
\begin{equation}
\frac{1}{2}\psi \rho^2+\kappa ^2\rho-3\phi H^2-3H\dot{\phi}+\frac{1}{2}V=0,
\end{equation}
which gives
\begin{equation}
\rho =\frac{\kappa ^2}{\psi}\left[-1\pm\sqrt{1+\frac{2\psi }{\kappa ^4}\left(3\phi H^2+3H\dot{\phi}-\frac{V}{2}\right)}\right].
\end{equation}
By series expanding the square root we obtain
\begin{equation}\label{49}
\rho \approx 6\phi \rho _m+\frac{2}{\kappa ^2}\left(3H\dot{\phi}-\frac{V}{2}\right),
\end{equation}
where we have denoted by $\rho_m=3H^2/\kappa ^2$ the general relativistic expression of the matter density. Eq.~(\ref{49}) shows that significant differences in the behavior of the matter density occur in the present model, as compared to standard general relativity. The two matter densities may become very close if $\phi \approx 6$, and $V\approx 6H\dot{\phi}$, or if $\phi \approx 6$, and $2\left(3H\dot{\phi}-V/2\right)/\kappa ^2<< \rho _m$.

\section{Cosmological Models}\label{sec:models}

In the present Section, we investigate the cosmological implications of $f\left(R,\mathbf{T}^2\right)$ gravity in the presence of matter creation. The dynamical evolution of the universe in this scenario is described by the generalized Friedmann equations (\ref{eq21}) and (\ref{eq22}), respectively, in the presence of a potential $V(\phi, \psi)$. Different forms of the potential may generate significantly different cosmological behaviors, and, in the following, we explore several models obtained by assuming different algebraic structures of $V$. In the following we restrict our analysis to flat cosmological models, with $k=0$, and to dust-like matter, with $p=0$.  As a first step, we consider the simple de Sitter type evolution of the model. Finally, more general cosmological models are investigated.

\subsection{The de Sitter solution: $H={\rm constant}$}

Currently, for a modified theory of gravity to be considered a viable alternative to GR, it must be able to accommodate the de Sitter solution. This solution possesses a constant Hubble function ($H = H_0 =$ constant) which can entail a late-time cosmic accelerated expansion.

\subsubsection{The vacuum solution}

In the case of vacuum, with $\rho =p=0$, and for $H=H_0={\rm constant}$, the potential $V$ takes the form $V(\phi,\psi)=12H_0^2\phi$. Equation (\ref{eq22}) becomes $H_0\dot{\phi}-\ddot{\phi}=0$, with the general solution
\begin{equation}
\phi(t)=C_1\frac{e^{H_0t}}{H_0}+C_2,
\end{equation}
where $C_1$ and $C_2$ are arbitrary constants of integration. Substituting these results into Eq.~(\ref{eq21}) gives $3C_2H_0^2=0$, which is identically satisfied for $C_2=0$. Hence, an exact vacuum de Sitter type solution can be found for an exponentially increasing $\phi =C_1 e^{H_0t}/H_0$, and for an arbitrary field $\psi$.
Note that the particle creation rate $\Gamma$, given by Eq.~(\ref{creation_rate}) is zero, $\Gamma \equiv 0$, for a vacuum universe, as expected.

\subsubsection{Constant energy density
}

Let us now assume that the matter density of the dust-like matter is constant, i.e., $p=0$ and $\rho=\rho_0$.
For this case, the self-interacting potential is obtained from Eqs. \eqref{eq23} and \eqref{eq24} by a simple integration, to yield
\begin{equation}
  V(\phi,\psi) = 12H_0^2\phi + \rho_0^2\psi + \Lambda_0,
  \label{vphipsi}
\end{equation}
where $\Lambda_0$ is the resulting constant of integration.

For this specific case, from Eq. \eqref{eq25} we obtain the differential equation
\begin{equation}
\dot\psi(t) + 3H_0\psi + \dfrac{3\kappa^2H_0 }{\rho_0}= 0,
\label{coc}
\end{equation}
which has the general solution
\begin{equation}
\psi(t)=e^{-3 H_0 t}\left[\psi_0+\dfrac{\kappa^2}{\rho_0}\left(1-e^{3 H_0 t}\right)\right],
\label{psi}
\end{equation}
where we have defined $\psi_0=\psi(0)$. On the other hand, if we take into consideration the Friedman equation in Eq. \eqref{eq21}, one can obtain the differential equation that describes the scalar field $\phi$ under a constant density solution and a flat spacetime, i.e. $k=0$, as
\begin{equation}
\dot\phi(t) -H_0\phi -\dfrac{1}{6H_0}\left[\psi\left(\rho_0^2-\rho_0\right)+2\kappa^2\rho_0+\Lambda_0\right] = 0,
\label{def_phi}
\end{equation}
which, taking into account the solution for $\psi$ in Eq. \eqref{psi}, leads us to the general solution
\begin{eqnarray}
    \phi(t) &=& \dfrac{1}{12 H_0^2}\Big\{e^{-3 H_0 t}\Big[-2 e^{3 H_0 t} \Lambda_0-\rho\left(k^2+\rho \psi_0\right)\nonumber\\
   && +e^{4 H_0 t}\left(2 \Lambda_0+\kappa^2 \rho+12 H_0^2 \phi_0+\rho^2 \psi_0\right)\Big]\Big\},
\end{eqnarray}
where $\phi_0=\phi(0)$. One can now obtain the creation pressure from Eq. \eqref{pc} and the creation rate from Eq. \eqref{creation_rate}, which can be simplified using Eq. \eqref{coc} yielding
\begin{equation}
\Gamma = 3H_0,\qquad p_c =-\rho_0.
\end{equation}

In the present de Sitter type model, the decrease in the matter density due to the cosmological expansion is exactly compensated by the matter creation processes, leading to a constant density Universe. By assuming for $H_0$ a value of the order of $H_0=2.27\times 10^{-18}$ s$^{-1}$ \cite{Planck:2018vyg}, we obtain for the particle creation rate the numerical value $\Gamma =6.81 \times 10^{-18}$ s$^{-1}$. The variation of the total particle number is given by $N(t)/N_0=\exp\left[3H_0\left(t-t_0\right)\right]$, with the particle number increasing during a time interval $t=1/H_0$, of the order of the age of the Universe, as $\left.N(t)\right|_{t=1/H_0}/N_0=e^3=19.903$. Hence, for a Universe in a de Sitter type expansion, in the presence of matter creation with $\Gamma =3H_0$ during its entire history, matter creation led to an increase of the initial particle number existing at the Big Bang by a factor of around 20.

\subsubsection{Time-dependent energy density}

In order to obtain a more realistic cosmological model, let us now analyze the scenario for which the matter density can also vary in time, so that $\rho=\rho(t)$. The de Sitter solution for a non-constant matter density requires a more careful analysis. For an exponentially expanding de Sitter universe with non-zero matter density the generalized Friedmann equations, Eqs. (\ref{eq21}) and (\ref{eq22}), take the following forms
\begin{equation}
3H_0^2\phi=\kappa ^2\rho +\frac{1}{2}V+\frac{1}{2}\psi \rho^2-3H_0\dot{\phi},
\end{equation}
\begin{equation}\label{eq52}
-\kappa ^2\rho-\psi \rho^2+H_0\dot{\phi}-\ddot{\phi}=0,
\end{equation}
respectively. Note that in this model the solution of the field equations, as well as the cosmological behavior, explicitly depends on the functional form of the potential $V$.  To obtain a simple toy model we assume that $V$ is given by
\begin{equation}\label{V1}
V(\phi,\psi)=12H_0^2\phi-\frac{\alpha ^2}{\psi} ,
\end{equation}
where $\alpha$ is a constant parameter, from which, by using Eq.~(\ref{eq24}), we obtain $V_\psi=\alpha^2 /\psi ^2=\rho ^2$, which fixes $\psi$ as $\psi =\pm \sqrt{\alpha ^2/\rho ^2}=-\alpha/\rho $, with $\alpha>0$, $\rho >0$, giving immediately $\dot{\psi}=\alpha \dot{\rho}/\rho^2$.

Equation~\eqref{eq25} takes the form
\begin{equation}
    \dot{\rho}\left(\kappa^2+\psi\rho\right) + \rho^2\left(\dot{\psi}+3H_0\psi\right)+3\kappa^2H_0\rho = 0,
\end{equation}
and, for our specific form of the potential, reduces to
\begin{equation}
\dot{\rho}+3H_0\left(1-\frac{\alpha}{\kappa ^2}\right)\rho=0,
\end{equation}
giving for the matter density variation the expression
\begin{equation}
\rho (t)=\rho _0e^{-3H_0\left(1-\frac{\alpha}{\kappa ^2}\right)t},
\end{equation}
where $\rho_0=\rho (0)$. Under these results, Eq.~(\ref{eq52}) becomes
\begin{equation}
\ddot{\phi}-H_0\dot{\phi}+\left(\kappa ^2-\alpha\right)\rho _0e^{-3H_0\left(1-\frac{\alpha}{\kappa ^2}\right)t}=0,
\end{equation}
with the general solution given by
\begin{equation}
\phi (t)=\frac{c_1 e^{H_0 t}}{H_0}+c_2-\frac{\kappa ^4 \rho_0 e^{-\frac{3 H_0 t
   \left(\kappa ^2-\alpha \right)}{\kappa ^2}}}{3 H_0^2 \left(4 k^2-3\alpha\right)},
   \label{de_sitter_t_phi}
\end{equation}
where $c_1$ and $c_2$ are arbitrary constants of integration. Substituting the obtained expressions for $\rho$, $\psi$, and $\phi$ into the first Friedmann equation in Eq. (\ref{eq21}) we obtain again $3H_0^2c_2=0$, which requires that $c_2=0$.

Consider now the particle creation rate in Eq. \eqref{creation_rate}. Under the results obtained in the previous paragraphs, we have
\begin{equation}
\Gamma =3\frac{\alpha}{\kappa ^2}H_0>0.
\end{equation}
Hence, from the point of view of the thermodynamics of the open systems, a de Sitter type solution corresponding to the potential in Eq. (\ref{V1}) would require the presence of a particle creation process, with a positive matter creation rate. By redefining $\alpha$ as $\alpha =n \kappa ^2$, $n>0$, the particle creation rate can be written as $\Gamma =3nH_0$, giving for the variation of the total particle number of the Universe the expression $N(t)/N_0=\exp\left[3nH_0\left(t-t_0\right)\right]$. The increase of the total particle number in a time interval of the order of the age of the Universe is $\left.N(t)\right|_{t=1/H_0}/N_0=e^{3n}$, a value which depends on the choice of the constant $\alpha$ in the potential. The time variation of the matter energy density, $\rho(t)/\rho _0=\exp \left[-3H_0(1-n)t\right]$, depends on the numerical value of $n$. For $n<1$, the energy density of matter is decreasing, while for $n>1$ it is increasing. For a time interval of the order of the age of the Universe $t=1/H_0$, the variation of the matter density due to matter creation is $\left.\rho(t)\right|_{t=1/H_0}/\rho _0=\exp \left[-3(1-n)\right]$. For example, for $n=2$, the increase in the baryonic matter density from the Big Bang up to the present time due to matter creation is  $\left.\rho(t)\right|_{t=1/H_0}/\rho _0=\exp \left[3)\right]=19.903$.

\subsection{Cosmological models with varying $H$}

We consider now general cosmological models, for which the Hubble function $H$ is a function of time, reaching a constant value at the end of a complex cosmological evolution.  In order to simplify the mathematical formalism, we introduce first the set of new variables $\left(h,\tau,r,U,\Psi\right)$, defined according to the relations
\begin{eqnarray}
H=H_0h, \quad \tau =H_0t, \quad  \rho=\frac{3H_0^2}{\kappa ^2}r,
    \nonumber \\
V=3H_0^2U, \quad \psi=\frac{\kappa ^4}{9H_0^2}\Psi.
    \nonumber
\end{eqnarray}
A strategy to find solutions is to note that the field equations, the scalar field equations, and the conservation equation, form a system of five equations, of which four are linearly independent. After one specifies the potential $U$, one obtains a determined system of four equations to four unknown functions.

The full set of equations describing the cosmological evolution in the scalar-tensor representation of $f\left(R,\mathbf{T}^2\right)$ gravity is given by the two field equations, the conservation equation, and the equations of motion for the scalar fields, which under the definitions above take the forms
\begin{equation}\label{F1}
h^2=\frac{1}{\phi}\left(r+\frac{1}{2}U+\frac{1}{6}\Psi r^2-h\frac{d\phi}{d\tau}\right),
\end{equation}
\begin{equation}\label{F2}
\frac{dh}{d\tau}=\frac{1}{\phi}\left(-\frac{3}{2}r-\frac{1}{2}\Psi r^2+\frac{1}{2}h\frac{d\phi}{d\tau}-\frac{1}{2}\frac{d^2\phi}{d\tau ^2}\right),
\end{equation}
 \begin{equation}\label{F3}
 \left(1+\frac{1}{3}r\Psi\right)\frac{dr}{d\tau}+3hr+\frac{1}{3}r^2\left(\frac{d\Psi}{d\tau}+3h\Psi\right)=0,
 \end{equation}
  \begin{eqnarray}
 U_\phi &=& 2\left(\frac{dh}{d\tau}+2h^2\right),
 \label{Uphi}
    \\
 U_\Psi &=&\frac{1}{3}r^2,
 \end{eqnarray}
respectively.
Now, one may also note that the system of equations above contains five equations of which only four are dynamically independent. Furthermore, the system contains a total of five unknown functions $(h,r,\phi,\Psi, U)$. Once a specific form of the function $U$ is given, the number of dynamically independent functions equals the number of unknown functions and the system is determined without the need for any additional constraints.

 Moreover, inserting Eqs. \eqref{F1} and \eqref{F2} into Eq. \eqref{Uphi}, we obtain
 \begin{equation}
 U_\phi=\frac{2}{\phi}\left(\frac{1}{2}r-\frac{1}{6}\Psi r^2-\frac{3}{2}h\frac{d\phi}{d\tau}-\frac{1}{2}\frac{d^2\phi}{d\tau^2}\right).
 \end{equation}

 To facilitate the comparison with the observational data we reformulate the evolution equations in terms of the redshift variable $z$, defined by
 \begin{equation}
 1+z=\frac{1}{a}.
 \end{equation}
 Hence we obtain
 \begin{equation}
 \frac{d}{d\tau}=-(1+z)h\frac{d}{dz}.
 \end{equation}

 In the redshift space, and upon the introduction of a new variable $d\phi/d\tau =\omega$, the equations describing the cosmological evolution can be recast as a first order dynamical system given by
 \begin{equation}\label{d1}
 -(1+z)h\frac{d\phi}{dz}=\omega,
 \end{equation}
 \begin{eqnarray}\label{d3}
 &&-(1+z)\left(1+\frac{1}{3}r\Psi\right)h\frac{dr}{dz}+3hr\nonumber\\
 &&\hspace{1cm}+\frac{1}{3}r^2\left[-(1+z)h\frac{d\Psi}{dz}+3h\Psi\right]=0,
 \end{eqnarray}
 \begin{equation}\label{d4}
 (1+z)h\frac{d\omega}{dz}=\phi U_\phi-r+\frac{1}{3}\Psi r^2+3h\omega,
 \end{equation}
 \begin{equation}\label{d5}
 (1+z)h\frac{dh}{dz}=2h^2-\frac{1}{2}U_\phi,
 \end{equation}
 \begin{equation}\label{d6}
  U_\Psi=\frac{1}{3}r^2.
 \end{equation}

 The system of Eqs. (\ref{d1})-(\ref{d6}) must be solved together with the initial conditions $h(0)=1$, $r(0)=r_0$, $\phi (0)=\phi_0$, and $\omega (0)=\omega_0$, respectively.

 In this framework and under the definitions and assumptions outlined above, the particle creation rate takes the form
\begin{equation}
\Gamma =H_0\left[3h-(1+z)\frac{1}{r}\frac{dr}{dz}\right]=H_0\Sigma,
\end{equation}
whereas the dimensionless creation pressure becomes
\begin{equation}
P_c=\frac{p_c\kappa ^2}{H_0^2}=-\frac{r}{h}\Sigma.
\end{equation}

 In order to describe the accelerating/decelerating nature of the dynamical evolution of the universe we introduce the deceleration parameter $q$, defined as
 \begin{equation}
 q=\frac{d}{d\tau}\frac{1}{h}-1=(1+z)\frac{1}{h}\frac{dh}{dz}-1.
 \end{equation}
 The negative sign of $q$ indicates an accelerated expansion of the universe, while a positive sign corresponds to decelerating phases of evolution. In the following we investigate two distinct cosmological models, corresponding to two choices of the potential $U$.

 To test the viability of our models,
 we compare them to the standard $\Lambda
$CDM model,  and with a specific set of observational data for the Hubble function.
The Hubble function of the $\Lambda $CDM model is given by
\begin{equation}
H=H_{0}\sqrt{\frac{\Omega _{m}^{(c)}}{a^{3}}+\Omega _{\Lambda }}=H_{0}\sqrt{%
\Omega _{m}^{(c)}(1+z)^{3}+\Omega _{\Lambda }},
\end{equation}%
where we have denoted $\Omega _{m}^{(c)}=\Omega _{b}^{(c)}+\Omega _{DM}^{(c)}$, while $%
\Omega _{b}^{(c)}=\rho _{b}/\rho _{cr}$, $\Omega _{DM}^{(c)}=\rho _{DM}/\rho
_{cr} $ and $\Omega _{\Lambda }=\Lambda /\rho _{cr}$, where $\rho_{cr}$ is the critical density of the universe,  represent the density
parameters of the baryonic matter, dark matter, and dark energy,
respectively. The deceleration parameter is given in the $\Lambda$CDM model
by the relation
\begin{equation}
q(z)=\frac{3(1+z)^{3}\Omega _{m}^{(c)}}{2\left[ \Omega _{\Lambda }+(1+z)^{3}\Omega
_{m}^{(c)}\right] }-1.
\end{equation}
For the matter and dark energy density parameters
of the $\Lambda $CDM model we adopt in the following the values $\Omega _{DM}^{(c)}=0.2589$, $\Omega
_{b}^{(c)}=0.0486$, and $\Omega _{\Lambda }=0.6911$, respectively \cite{Planck:2018vyg}. Hence, it follows that the
total matter density parameter $\Omega _{m}^{(c)}=\Omega _{DM}^{(c)}+\Omega _{b}^{(c)}=0.3089$. The present day value of the
deceleration parameter predicted by the $\Lambda$CDM model can thus be obtained as  $q(0)=-0.5381$, showing that the present day universe
is in an accelerating phase.

\subsubsection{Model I: $U(\phi,\Psi)=\alpha \phi^{n+1}+\frac{1}{3}\beta \Psi ^{m+1}$}

As a first example of a cosmological model we consider the case in which the potential $U(\phi, \psi)$ has a simple additive algebraic structure, so that
 \begin{equation}
 U(\phi,\Psi)=2\alpha \phi^{n+1}+\frac{1}{3}\beta \Psi ^{m+1},
 \end{equation}
 where $\alpha$, $\beta$, $n$ and $m$ are constants defined in such a way for later convenience. Equation~(\ref{d6}) gives immediately
 \begin{equation}
 \Psi=\left[\frac{r^2}{\beta(m+1)}\right]^{1/m}.
 \end{equation}

 In this case the cosmological evolution equations take the form
  \begin{equation}\label{m1a}
 -(1+z)h\frac{d\phi}{dz}=\omega,
 \end{equation}
 \begin{equation}\label{m1b}
 (1+z)h\frac{dh}{dz}=2h^2-\alpha (n+1)\phi ^n,
 \end{equation}
 \begin{eqnarray}\label{m1c}
 (1+z)h\frac{d\omega}{dz}&=&2\alpha (n+1)\phi ^{n+1} -r
    \nonumber\\
 && \hspace{-1cm} +\frac{1}{3}\beta (m+1)\left[\frac{r^2}{\beta(m+1)}\right]^{1/m+1}+3h\omega,
 \end{eqnarray}
 \begin{eqnarray}\label{m1d}
&& (1+z)\left\{1+\frac{1}{3}r\left[\frac{r^2}{\beta(m+1)}\right]^{1/m}\right\}h\frac{dr}{dz} - 3hr
 - \frac{1}{3}r^2 \times
    \nonumber\\
&&\hspace{0.35cm} \times \left[-(1+z)h\frac{2}{\beta m(m+1)}r\Psi^{1-m}\frac{dr}{dz}+3h\Psi\right]=0.
 \end{eqnarray}

\begin{figure*}[htbp]
\centering
\includegraphics[width=8.0cm]{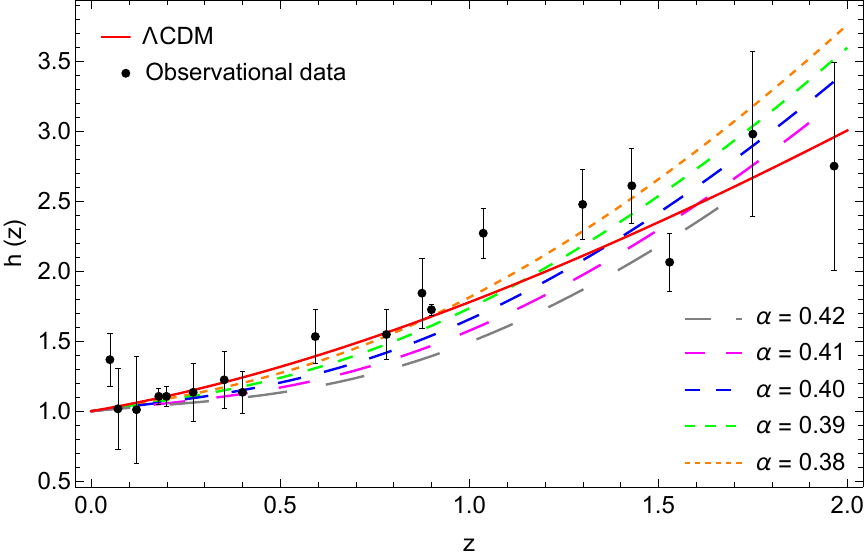} %
\includegraphics[width=8.0cm]{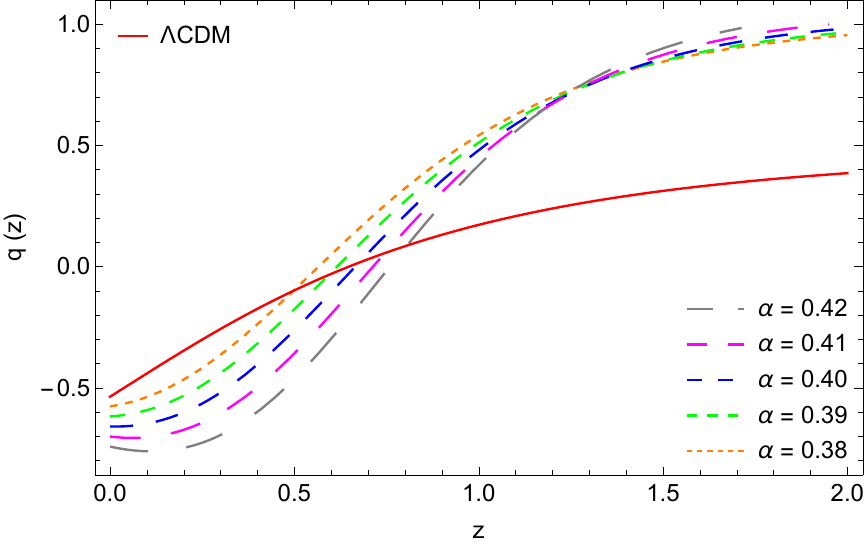}
\caption{Variation of the dimensionless Hubble function $h(z)$ (left panel),
and of the deceleration parameter $q(z)$ (right panel), in the scalar-tensor representation of the $f\left(R,\mathbf{T}^2\right)$ cosmological model
with $U(\phi,\Psi)=\alpha \phi^{n+1}+\frac{1}{3}\beta \Psi ^{m+1}$, for $n=1$, $m=2$, $\beta =1$, and for different values of $\alpha$: $\alpha =0.38$ (dotted
curve), $\alpha =0.39$ (short dashed curve), $\alpha =0.40$ (dashed curve), $\alpha =0.41$ (long dashed curve), and $\alpha =0.42$ (ultra-long dashed curve),
respectively. The system of the cosmological evolution equations was integrated numerically with the initial conditions $h(0)=1$, $r(0)=0.03$, $\phi (0)=2.1$, and $\theta (0)=-1.5$. The variations of the Hubble function and of the deceleration parameter of the $\Lambda$CDM model are represented by
the red solid line. The observational data are shown together with their error bars \cite{Moresco:2015cya,Boumaza:2019rpt}.}
\label{fig1}
\end{figure*}

 The system of differential equations in Eqs. (\ref{m1a})-(\ref{m1d}) must be solved with the initial conditions $h(0)=1$, $r(0)=r_0$, $\phi (0)=\phi_0$, and $\omega (0)=\omega_0$, respectively. The variations with respect to the redshift of the Hubble function and of the deceleration parameter are represented in Fig.~\ref{fig1}. The cosmological evolution is strongly dependent on the model parameters, as well as on the initial conditions for the scalar field $\phi$, and its first-order time derivative. The model provides results in a close agreement with the observational data for the Hubble function up to a redshift $z\approx 1$, but at higher redshifts significant differences appear, at least for the considered values of the model parameters. Generally, at higher redshifts $z>1$, the expansion rate of the universe increases faster as compared to standard cosmology. The differences in the evolution of the deceleration parameters of the two models are more significant, with the deceleration parameter of $f\left(R,\mathbf{T}^2\right)$ gravity taking higher values, close to $q\approx 1$, at redshifts $z>1.5$. The $f\left(R,\mathbf{T}^2\right)$ model also predicts a transition to an accelerated expansion state at redshifts $z \lesssim 0.5$, but with the present day value of the deceleration parameter strongly dependent on the model parameters.

\begin{figure*}[htbp]
\centering
\includegraphics[width=8.0cm]{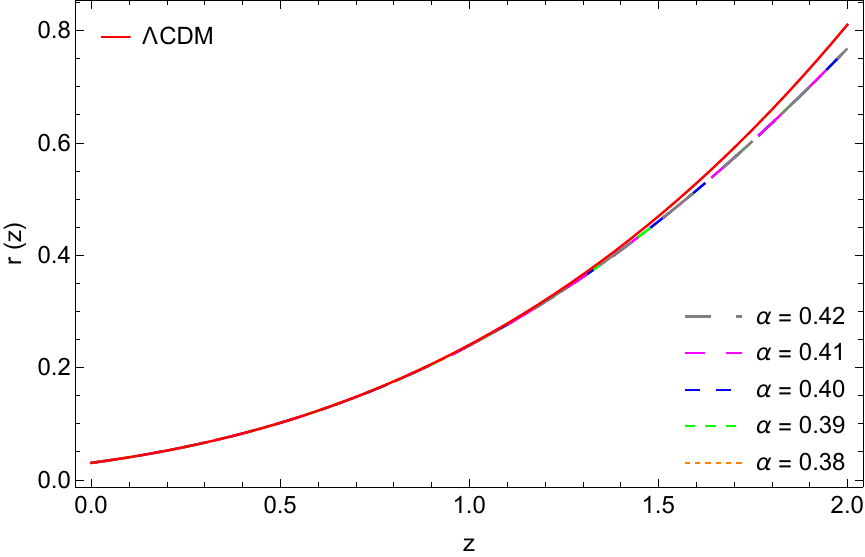} %
\includegraphics[width=8.0cm]{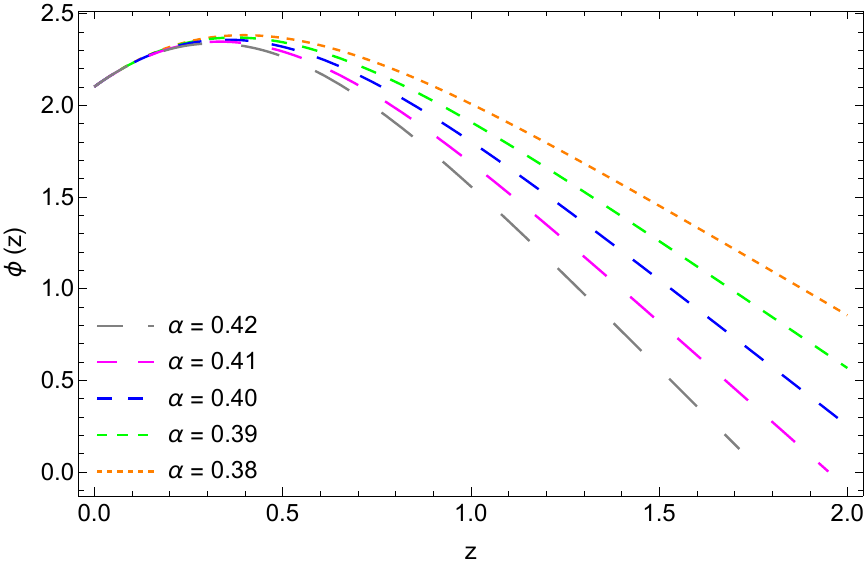}
\caption{Variation of the dimensionless matter density $r(z)$ (left panel),
and of the scalar field $\phi(z)$ (right panel), in the scalar-tensor representation of the $f\left(R,\mathbf{T}^2\right)$ cosmological model
with $U(\phi,\Psi)=\alpha \phi^{n+1}+\frac{1}{3}\beta \Psi ^{m+1}$, for $n=1$, $m=2$, $\beta =1$, and for different values of $\alpha$: $\alpha =0.38$ (dotted
curve), $\alpha =0.39$ (short dashed curve), $\alpha =0.40$ (dashed curve), $\alpha =0.41$ (long dashed curve), and $\alpha =0.42$ (ultra-long dashed curve),
respectively. The system of the cosmological evolution equations was integrated numerically with the initial conditions $h(0)=1$, $r(0)=0.03$, $\phi (0)=2.1$, and $\theta (0)=-1.5$.
The variations of the dimensionless matter density of the $\Lambda$CDM model are represented by
the red solid line.}
\label{fig2}
\end{figure*}

The variations of the ordinary (baryonic)  matter energy density and of the scalar field $\phi$ are represented in Fig.~\ref{fig2}. The prediction of the matter density evolution of our model  basically coincides with that of the $\Lambda$CDM. Furthermore, the matter distribution does not depend significantly on the model parameters, however, some differences may appear. This coincidence indicates that indeed the extra terms coming from the contribution of the two scalar fields could be interpreted as dark energy, and they trigger the accelerated cosmological expansion of the universe. The scalar field $\phi$ has a complicated evolution, initially increasing at small redshifts, and reaching a maximum at $z\approx 0.5$, followed by a decrease as $z$ increases. At large $z$ values, the behavior of $\phi$ is strongly dependent on the small variations of the model parameter $\alpha$.

\begin{figure*}[htbp]
\centering
\includegraphics[width=8.0cm]{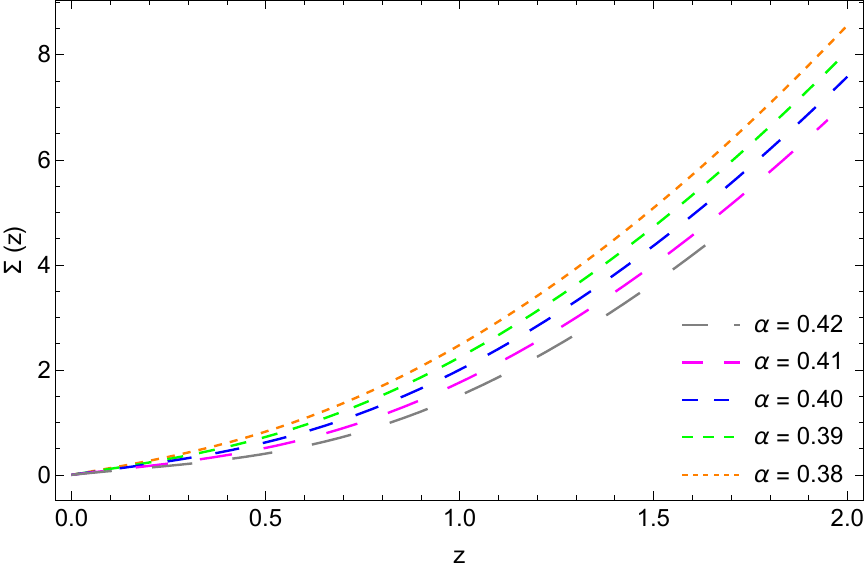} %
\includegraphics[width=8.0cm]{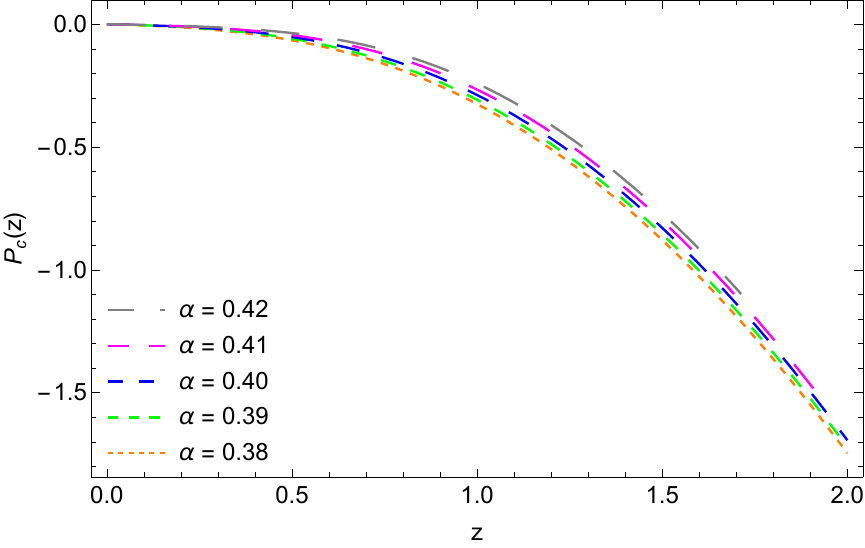}
\caption{Variation of the dimensionless particle creation rate $\Sigma (z)$ (left panel),
and of the dimensionless creation pressure  $P_c(z)$ (right panel), in the scalar-tensor representation of the $f\left(R,\mathbf{T}^2\right)$ cosmological model
with $U(\phi,\Psi)=\alpha \phi^{n+1}+\frac{1}{3}\beta \Psi ^{m+1}$, for $n=1$, $m=2$, $\beta =1$, and for different values of $\alpha$: $\alpha =0.38$ (dotted
curve), $\alpha =0.39$ (short dashed curve), $\alpha =0.40$ (dashed curve), $\alpha =0.41$ (long dashed curve), and $\alpha =0.42$ (ultra-long dashed curve),
respectively. The system of the cosmological evolution equations was integrated numerically with the initial conditions $h(0)=1$, $r(0)=0.03$, $\phi (0)=2.1$, and $\theta (0)=-1.5$. }
\label{fig3}
\end{figure*}

The redshift dependencies of the dimensionless particle creation rate $\Sigma$ and of the dimensionless creation pressure $P_c$ are shown in Fig.~\ref{fig3}. The particle creation rate monotonically increases as a function of the redshift (decreases with respect to time). Its evolution strongly depends on the model parameters, with the differences increasing at higher redshifts. The creation pressure takes negative values for the entire considered redshift range, indicating that the considered model satisfies the basic requirements of the thermodynamic of irreversible processes. The creation pressure has a lesser dependence on the model parameters, but it rapidly decreases at higher redshifts.

\subsubsection{Model II: $U(\phi,\Psi)=\alpha \phi +\frac{\beta}{3} \Psi+\frac{\gamma}{3} \phi ^n \Psi ^{m+1}$}

As a second example of a cosmological model we consider that the potential $U$ has a non-additive structure, represented as
 \begin{equation}
  U(\phi,\Psi)=\alpha \phi +\frac{1}{3}\beta \Psi+\frac{\gamma }{3}\gamma \phi ^n \Psi ^{m+1},
 \end{equation}
where $\alpha$, $\beta$, $\gamma$, $n$ and $m$ are constants defined in such a way for later convenience. Then we immediately find
\begin{equation}
U_\phi=\alpha +\frac{1}{3}\gamma n\phi ^{n-1}\Psi ^{m+1},
\end{equation}
\begin{equation}
U_\Psi=\frac{1}{3}\left[\beta +\gamma (m+1)\phi ^n\Psi ^{m}\right]=\frac{1}{3}r^2,
\end{equation}
from which we obtain
\begin{equation}
\Psi =\left[\frac{r^2-\beta}{(m+1)\gamma}\right]^{1/m}\phi ^{-n/(m)}.
\end{equation}

Hence, the system of the cosmological evolution equations for this model take the form
\begin{equation}\label{m2a}
 -(1+z)h\frac{d\phi}{dz}=\omega,
 \end{equation}
 \begin{equation}\label{m2b}
 (1+z)h\frac{dh}{dz}=2h^2-\frac{1}{2}\left[\alpha +\frac{1}{3}\gamma n\phi ^{n-1}\Psi ^{m+1}\right],
 \end{equation}
 \begin{eqnarray}\label{m2c}
 (1+z)h\frac{d\omega}{dz}&=&\alpha \phi +\frac{1}{3}\gamma n\phi ^{n}\Psi ^{m+1}-r\nonumber\\
 && \hspace{-0.5cm} +\frac{1}{3}\left[\frac{r^2-\beta}{(m+1)\gamma}\right]^{\frac{1}{m}}\phi ^{-\frac{n}{m}}r^2+3h\omega,
 \end{eqnarray}
 \begin{eqnarray}\label{m2d}
 &&-(1+z)\left\{1+\frac{1}{3}r\left[\frac{r^2-\beta}{(m+1)\gamma}\right]^{1/(m)}\phi ^{-n/m}\right\}h\frac{dr}{dz}
    \nonumber\\
 && \hspace{1.0cm} +3hr+\frac{1}{3}\left[\frac{r^2-\beta}{(m+1)\gamma}\right]^{1/(m)}\phi ^{-n/m}r^2 \times
    \nonumber\\
 &&\times \left[-\frac{2(1+z)}m\frac{hr}{r^2-\beta}\frac{dr}{dz}+\frac{n(1+z)}m\frac{h\omega}{\phi}+3\right]=0.
 \end{eqnarray}

\begin{figure*}[htbp]
\centering
\includegraphics[width=8.0cm]{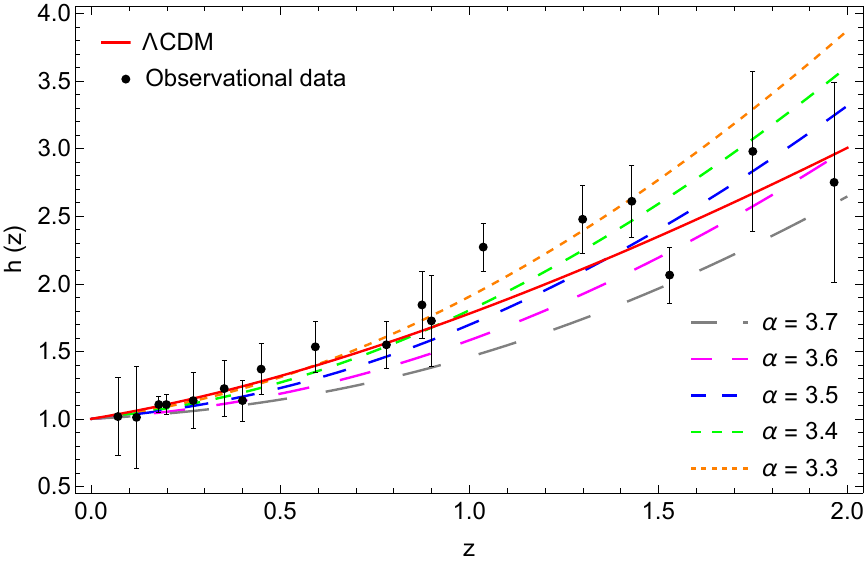} %
\includegraphics[width=8.0cm]{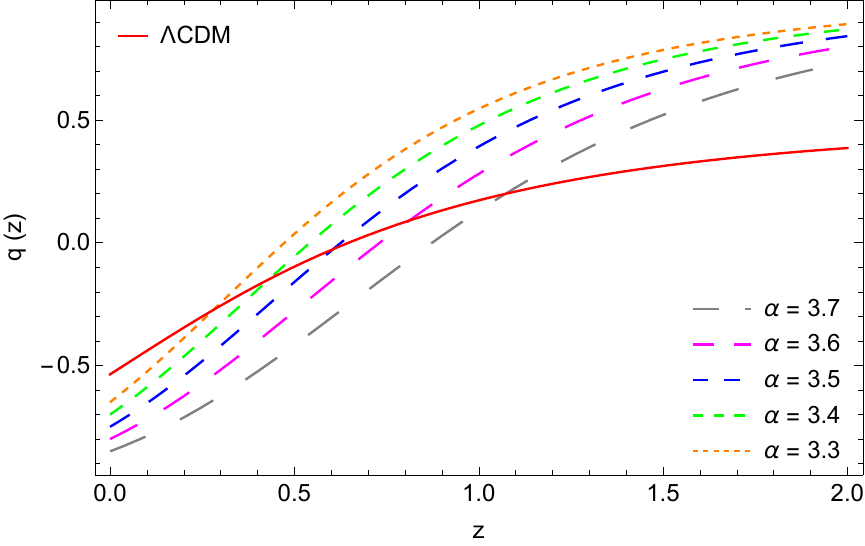}
\caption{Variation of the dimensionless Hubble function $h(z)$ (left panel),
and of the deceleration parameter $q(z)$ (right panel), in the scalar-tensor representation of the $f\left(R,\mathbf{T}^2\right)$ cosmological model
with $U(\phi,\Psi)=\alpha \phi +\frac{\beta}{3} \Psi+\frac{\gamma}{3} \phi ^n \Psi ^{m+1}$, for $n=10^{-20}$, $m=1$, $\beta =0.01$, $\gamma =0.004$, and for different values of $\alpha$: $\alpha =3.3$ (dotted curve), $\alpha =3.4$ (short dashed curve), $\alpha =3.5$ (dashed curve), $\alpha =3.6$ (long dashed curve), and $\alpha =3.7$ (ultra-long dashed curve), respectively. The system of the cosmological evolution equations was integrated numerically with the initial conditions $h(0)=1$, $r(0)=0.03$, $\phi (0)=0.001$, and $\theta (0)=-0.065$. The variations of the Hubble function and of the deceleration parameter of the $\Lambda$CDM model are represented by
the red solid line. The observational data are shown together with their error bars \cite{Moresco:2015cya,Boumaza:2019rpt}.}
\label{fig4}
\end{figure*}

The variations with respect to the redshift of the Hubble function and of the deceleration parameter are presented, for this model, in Fig.~\ref{fig4}, for fixed values of the model parameters $n$, $m$, $\beta$ and $\gamma $, respectively. For specific values of the parameters, the model can reproduce the behavior of the Hubble function of the $\Lambda$CDM model, and describes well the observational data. However, significant differences with respect to $\Lambda$CDM appear for the deceleration parameter. Even though qualitatively this $f\left(R,\mathbf{T}^2\right)$ type model describes the transition from a decelerating to an accelerating state, the high redshift and the low redshift behaviors are very different.

\begin{figure*}[htbp]
\centering
\includegraphics[width=8.0cm]{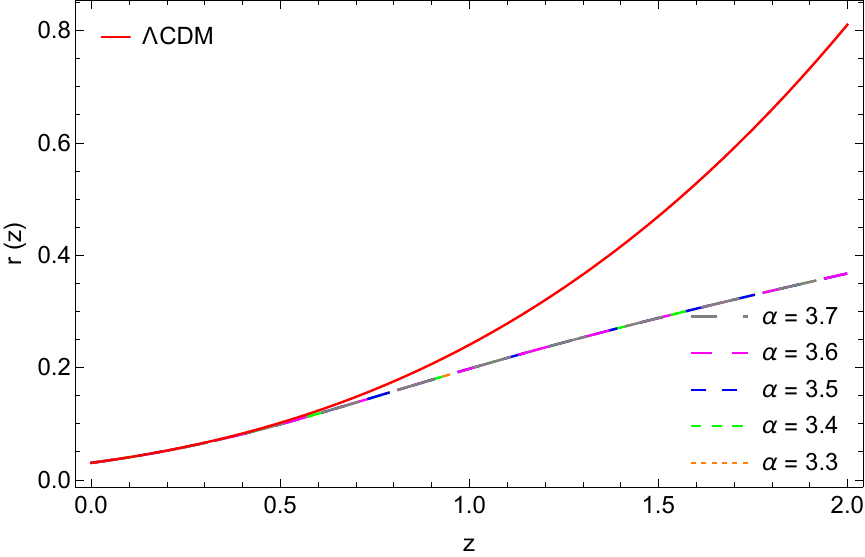} %
\includegraphics[width=8.0cm]{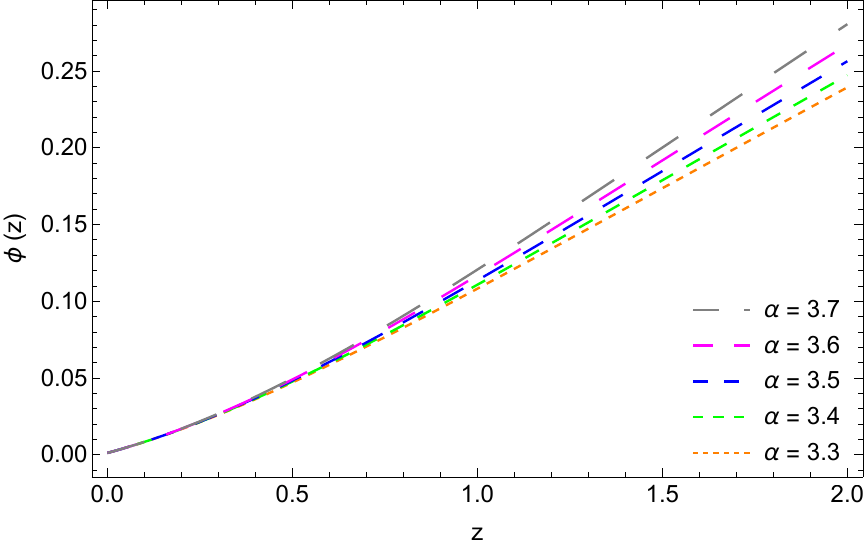}
\caption{Variation of the dimensionless matter density $r(z)$ (left panel),
and of the scalar field $\phi(z)$ (right panel), in the scalar-tensor representation of the $f\left(R,\mathbf{T}^2\right)$ cosmological model
with $U(\phi,\Psi)=\alpha \phi +\frac{\beta}{3} \Psi+\frac{\gamma}{3} \phi ^n \Psi ^{m+1}$, for $n=10^{-20}$, $m=1$, $\beta =0.01$, $\gamma =0.004$, and for different values of $\alpha$: $\alpha =3.3$ (dotted curve), $\alpha =3.4$ (short dashed curve), $\alpha =3.5$ (dashed curve), $\alpha =3.6$ (long dashed curve), and $\alpha =3.7$ (ultra-long dashed curve), respectively. The system of the cosmological evolution equations was integrated numerically with the initial conditions $h(0)=1$, $r(0)=0.03$, $\phi (0)=0.001$, and $\theta (0)=-0.065$. The variations of the dimensionless matter density of the $\Lambda$CDM model are represented by the red solid line.}
\label{fig5}
\end{figure*}

The variations of the matter densities in the $\Lambda$CDM and the cosmological model under study are presented in Fig.~\ref{fig5}. Interestingly enough, the matter densities coincide for $z\in (0,0.6)$, but for redshifts $z>0.6$ the matter content of the $\Lambda$CDM model increases much faster than that of the considered two scalar field model. This implies that at higher redshifts most of the energy content of the universe is in the form of dark energy, which corresponds to an effective energy generated by the two scalar fields, and of the two-field potential. Moreover, the matter density is basically independent on the variations of the model parameters.  This behavior can be understood with the help of Eq.~(\ref{49}). The function $\phi$ is increasing very slowly, while the term $V/2$, entering with the negative sign in the  expression of the density, determines its decrease at higher redshifts, as compared to the predictions of general relativity. Since the potential contains the term $\alpha \phi$, this term combines with the term $6\phi  \rho_m$, leading to a term of the form $(6-\alpha)\rho _m$. Since all the considered values of $\alpha$ satisfy the condition $(6-\alpha)\approx {\rm constant}$, the variation of the matter density is basically independent on the numerical values of $\alpha$.

The variation of the scalar field $\phi$ shows a monotonically, almost linear increase with the redshift (decrease in the cosmological time), and at redshifts $z>1$ its behavior depends on the considered model parameter $\alpha$.

\begin{figure*}[htbp]
\centering
\includegraphics[width=8.0cm]{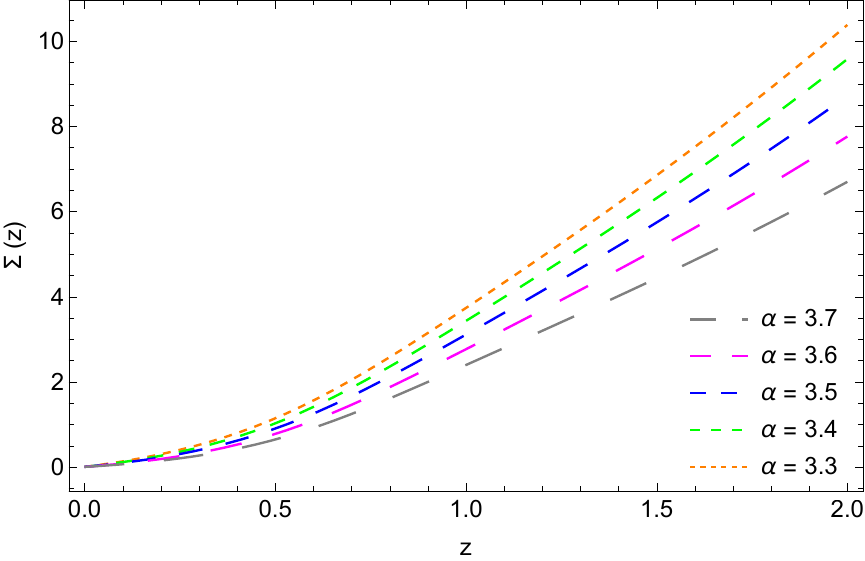} %
\includegraphics[width=8.0cm]{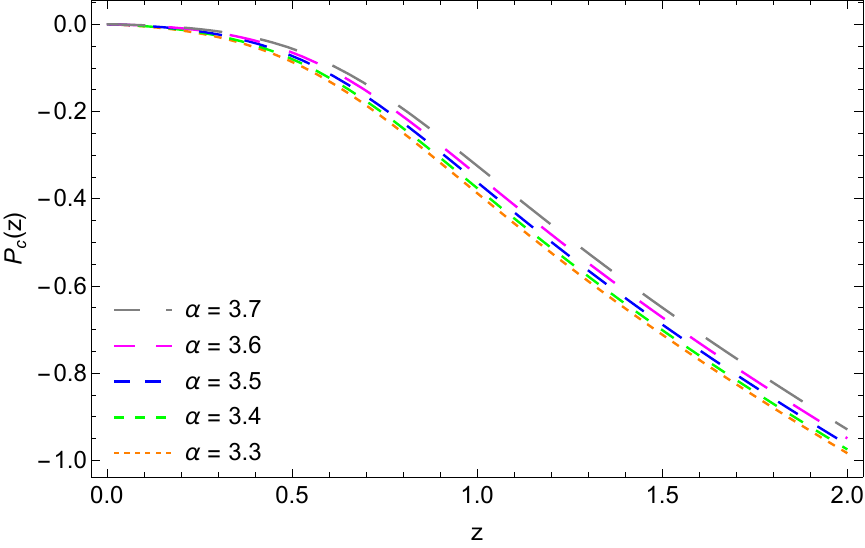}
\caption{Variation of the dimensionless particle creation rate $\Sigma (z)$ (left panel),
and of the dimensionless creation pressure  $P_c(z)$ (right panel), in the scalar-tensor representation of the $f\left(R,\mathbf{T}^2\right)$ cosmological model
with $U(\phi,\Psi)=\alpha \phi +\frac{\beta}{3} \Psi+\frac{\gamma}{3} \phi ^n \Psi ^{m+1}$, for $n=10^{-20}$, $m=1$, $\beta =0.01$, $\gamma =0.004$, and for different values of $\alpha$: $\alpha =3.3$ (dotted curve), $\alpha =3.4$ (short dashed curve), $\alpha =3.5$ (dashed curve), $\alpha =3.6$ (long dashed curve), and $\alpha =3.7$ (ultra-long dashed curve), respectively. The system of the cosmological evolution equations was integrated numerically with the initial conditions $h(0)=1$, $r(0)=0.03$, $\phi (0)=0.001$, and $\theta (0)=-0.065$. }
\label{fig6}
\end{figure*}

Finally, in Fig.~\ref{fig6}, the variations of the particle creation rate and of the creation pressure are represented as functions of the redshift. The particle creation rate is positive, and monotonically increases with the redshift (monotonically decreases in time), indicating that indeed the effective particle creation is the main factor triggering the accelerated expansion of the universe. The creation rate depends strongly on the model parameters even at low redshifts, and this dependence is stronger for large values of $z$. The creation pressure is negative, as required by the thermodynamic interpretation of the model, and it does not show a significant dependence on the model parameters.

\section{Summary and Discussion}\label{sec:concl}

In the present paper we investigated the cosmological implications of $f\left(R,\mathbf{T}^2\right)$ gravity, which belongs to a class of geometry-matter nonminimal coupling theories. Similarly to other types of theories with such couplings, the $f\left(R,\mathbf{T}^2\right)$ theory can also be formulated as an effective scalar-tensor theory, in terms of two scalar fields $\phi$ and $\psi$, and of an effective field potential $V(\phi, \psi)$. In the gravitational action the field $\phi$ couples directly to the Ricci scalar $R$, the field $\psi$ couples to the square of the matter energy-momentum tensor, while the potential $V(\phi,\psi)$ plays the role of a self-interaction potential for the two fields. The potential is not arbitrary, but it is determined, via its derivatives, by the scalar curvature and the square of the matter energy-momentum tensor. The field equations of the theory were obtained by varying the action with respect to the metric, and lead to a generalized set of Einstein-type field equations. Along with the equations of motion of the two scalar fields defined in this framework, one obtains a determined system of equations that can be solved for different forms of the interaction potential.

An interesting and important characteristic of this, and of similar theories with geometry-matter coupling, is the non-conservation of the standard matter energy-momentum tensor. In the present theory, $\nabla_\mu T^{\mu \nu}$ has a complicated mathematical structure, indicating a dependence on both scalar fields, and on the basic geometric and physical quantities. The theoretical interpretation of this result may suggest, at first sight, a pathological behavior of the theory. However, once a physical interpretation of this effect is found, it opens the way for a deeper understanding between various physical aspects that could play simultaneously an essential role in the description of the gravitational processes. In this work we have assumed that the nonconservation of the matter energy-momentum tensor can be interpreted physically in the framework of the thermodynamics of open systems, as considered for the first time in \cite{Prigogine:1988, Prigogine:1989zz, Prigogine:1986}. In this interpretation the non-conservation of the matter energy-momentum tensor describes the irreversible matter creation of matter by geometry, or equivalently the irreversible energy transfer from gravity to matter.

In the present thermodynamic formalism the basic quantity describing the deviations from GR is the particle creation rate $\Gamma$. When $\Gamma =0$, we recover all the results of standard cosmology. However, a nonzero particle creation rate, and the associated particle production, could offer new insights into some important astrophysical problems. One of the important challenges present day cosmology faces is related to the problem of dark matter. It is usually assumed that dark matter consists of weakly interacting, massive particles (WIMPS). Since no direct detection of dark matter particles has been yet recorded, it follows that they interact very weakly with baryonic matter particles. But this physical property makes the formation of dark matter in the early universe very difficult \cite{Her}, since dark matter couples only through the gravitational force, which is much weaker than the other interactions of nature. A promising solution to these difficulties is the gravitational particle creation of the dark matter. One can thus assume that the dark matter particles existing presently were created by the expansion of the Universe after the end of inflation by the energy transfer from the gravitational field to matter \cite{Her}. Hence, the present thermodynamic formalism as used for the interpretation of modified gravity theories with geometry-matter coupling could provide an effective way for explaining the physical properties of dark matter, and of the origin of dark matter particles.

In summary, in this work, we investigated in detail the cosmological evolution of the universe in the two field representation of $f\left(R,\mathbf{T}^2\right)$ gravity. Firstly, we considered in detail the possibility of the existence of a de Sitter solution in this model, where we analysed exponentially expanding solutions for the vacuum case, for a constant density universe, as well as for a time varying matter density. In all these three cases we explicitly showed that a de Sitter type solution exists. From the point of view of the particle production rate, the vacuum case corresponds to $\Gamma =0$, the constant density case requires $\Gamma =3H_0>0$, while for the time varying density case we obtain $\Gamma =3\alpha H_0/\kappa ^2>0$. This result was obtained for the case of a simple toy model, with the potential chosen, for mathematical convenience, as given by Eq.~(\ref{V1}).

Let us consider again the definition of the particle creation rate, as given by Eq.~(\ref{creation_rate}), which for $p=0$ can be written as
\begin{equation}
\Gamma =-\frac{1}{\kappa ^2}\left[\left(\dot{\psi}+3H\psi\right)\rho+\psi \dot{\rho}\right]=-\frac{1}{\kappa ^2}\left[\frac{d}{dt}\left(\psi \rho\right)+3H\psi \rho\right].
\end{equation}
 If the function $\psi$ satisfies the condition $\psi \rho ={\rm constant}=-\alpha$, then the creation rate is obtained as $\Gamma =3\alpha H/\kappa ^2$, and it is fully determined by the Hubble function describing the expansion of the Universe.

 On the other hand,  in the present particle creation model, by considering the zero pressure case, the particle creation rate is given by $\Gamma =3H+\dot{\rho}/\rho\geq 0$.  By using this relation in the particle balance equation, see Eq. (\ref{cov_nd}), we obtain that the newly created particle obeys the relation $\rho =kn$, where $k$ is a constant. From the first Friedmann equation it follows, in the general relativistic approximation, that $\rho \propto H^2$, which also gives $n\propto H^2$, results that are qualitatively similar to the estimations of the quantum field theory in curved spacetime. However, in the present model, other correction terms to the particle number density, and energy density appear.

 It is also interesting to note that the expression of the particle creation rate imposes string restrictions on the matter density evolution in the de Sitter phase, with $H=H_0={\rm constant}$. Particle creation stops if the density decreases as $\rho (t)=\rho_0 \exp\left[-3H_0\left(t-t_0\right)\right]$. A much more rapid decay of the density, of the form, say $\rho (t)=\rho_0 \exp\left[-4H_0\left(t-t_0\right)\right]$, is forbidden by the condition of the positivity of $\Gamma$. Generally, particle creation can take place in a de Sitter phase only if the exponentially decaying matter density is of the form $\rho (t)=\rho_0 \exp\left[-nH_0\left(t-t_0\right)\right]$, with $n>3$.

We also considered in our analysis theoretical cosmological models that may be used to test the viability of the considered modified gravity model. We investigated two distinct classes of models, obtained by two different choices of the two field scalar potential $V$. In order to facilitate the comparison with the cosmological observations, we reformulated the generalized Friedmann equations in the redshift space. A comparison with the predictions of the $\Lambda$CDM model, and with the observational data was also performed. We point out that for comparison with the determinations of the Hubble function we used a trial and error method, and no fitting was used in our analysis. Of course such an ``empirical'' method has its limitations, and only a full fitting of the existing observational datasets can give a clear understanding of the viability of this modified gravity theory. However, our preliminary results point towards the possibility that the considered models could not only give a qualitative, but also a quantitative description of the observational data, once the optimal values of the model parameters are obtained. Both our models give at least a qualitative description of a small set of the observational Hubble data. However, important quantitative differences with respect to the $\Lambda$CDM predictions appear in the case of the deceleration parameter. Since no direct observational results for a large redshift range exist for this important cosmological quantity, the behavior of $q$ in the considered two cosmological models cannot automatically rule out their viability. A proper fitting of the observational data may also improve the consistency with the $\Lambda$CDM predictions.

 We would like to point out that the thermodynamical formalism used in the present paper to describe matter creation can be easily extended to other modified gravity theories involving nonminimal couplings between matter and geometry. An interesting example in this direction may be theories involving a nonminimal coupling between matter and the Einstein tensor, with action given by \cite{Saridakis1}
\begin{eqnarray}
S &=& \int \sqrt{-g}\, d^4x
 \Big\{
 \frac{1}{2\kappa^2}\left(R-2\Lambda\right)
 \nonumber\\
 && +G_{\mu\nu}\left[\alpha
T^{\mu\nu}+\beta
(\partial^{\mu}T)(\partial^{\nu}T)\right]+ \mathcal{L}_m
 \Big\},
\end{eqnarray}
where $ G_{\mu\nu}$ is the Einstein tensor, $T_{\mu \nu}$ is the matter energy-momentum tensor, $T$ is its trace, and $\alpha $ and $\beta$ are constants. The matter conservation energy for this model takes the form
\begin{align}\label{nonconssar}
\nabla^\mu \left[
T_{\mu\nu}+\alpha T_{\mu\nu}^{(\alpha)}+\beta
T_{\mu\nu}^{(\beta)}\right]= 0.
\end{align}
For the definition of the quantities $T_{\mu\nu}^{(\alpha)}$ and $T_{\mu\nu}^{(\beta)}$ see \cite{Saridakis1}. Eq.~(\ref{nonconssar}) can be interpreted as describing irreversible matter creation in the model,

In the numerical investigations of our models we have adopted for $H_0$, the present day value of the Hubble function, the standard Planck value, $H_0=\left(67.4\pm 0.5\right)$ km/s/Mpc \cite{Planck:2018vyg},  different from the value $H_0=73\pm 1$ km/s/Mpc, obtained from the SH0ES team calibrating Type Ia Supernovae (SNIa) with Cepheids \cite{Valentino}. In our numerical analyisis, we have normalized the Hubble function values by using the Planck value for $H_0$. An alternative normalization of the observational data by using $H_0$ from the supernovae data is certainly possible, but it will give only minor modifications in the numerical values of the normalized Hubble function. On the other hand, we would like to point out that our main goal in the present work is to investigate the validity of the considered theoretical model for describing the cosmological data, at least at a qualitative level. Therefore, the change from one value (Planck) of $H_0$ to the other value (supernovae) will not modify our results in an essential way.

 However, in the present model it is possible to at least alleviate the Hubble tension. As one can see from the first Friedmann equation Eq.~(\ref{eq21}), the present day value of the Hubble function is determined  by the present day values of the matter energy density, of the two scalar fields, and of the potential, by the relation
  \begin{equation}
  H_0^2\frac{\dot{\phi}(0)}{\phi(0)}H_0=\kappa ^2\frac{\rho (0)}{3\phi (0)}+\frac{1}{6}\frac{V(0)}{\phi (0)}+\frac{1}{6}\frac{\psi (0)}{\phi (0)}\rho ^2 (0).
  \end{equation}

 Hence, by appropriately choosing the initial values of the fields $\phi$ and $\psi$, of their derivatives, and of the potential $V$, one could obtain concordance with the supernova data.

 On the other hand, the function $H(z)$ is generally given by
 \begin{equation}
 H(z)=H_0 A(z).
 \end{equation}
 where
 \begin{equation}
 A(z)=\sqrt{\frac{r(z)+U(z)/2+\Psi (z)r^2(z)}{1-(1+z)\frac{d}{dz}\ln \phi (z)}}.
 \end{equation}

If the $\Lambda$CDM model is correct for all redshifts, we should have $H(0)=H_0A(0)$, $A(0) = 1$, and we could fix $H_0$ as taking the value following from the supernova determinations, $H_0=H_0^{sup}=H(0)/A(0)$. If the $\Lambda$CDM model is not correct, then $A(0)$ could be different from one, or of the order of one, with small deviations from $\Lambda$CDM. Estimated from the early Universe, $A(z)$ could give a different value for the value of $H_0$. Let us assume that at a certain redshift $z_r$ $A\left(z_r\right)=0.92$. Then $H_0=H\left(z_r\right)/A\left(z_r\right)=H_0^{Planck}$. Thus we obtain
\begin{equation}
\frac{H_0^{Planck}}{H_0^{sup}}=\frac{H\left(z_r\right)}{H(0)}\times \frac{A(0)}{A\left(z_r\right)}=\zeta.
\end{equation}

Assuming $\zeta =0.92$, we obtain $H_0^{Planck}=\zeta\times H_0^{sup}=0.92\time 73=67.16$ km/s/Mpc. Such a numerical value for $\zeta$ could be obtained in the present model once the initial conditions, as well as the form of the potential $U$ are known. We can also consider as an additional requirement the condition of the  small deviations from the $\Lambda$CDM model, especially during the period of formation of the galaxies and stars, that is, from the reionization phase, or from the cosmic dawn era, $z <11$.

The behavior of the cosmological models also essentially depends on the choice of the two scalar field potential $V$. In our analysis this choice is relatively arbitrary, and is suggested by the necessity of explaining phenomenologically the observational data. Several constraints on the potential can be obtained from other astrophysical and cosmological phenomena, such as black hole solutions, gravitational lensing, structure formation, or from the study of the gravitational waves.

To conclude, $f\left(R,\mathbf{T}^2\right)$ gravity represents an interesting theoretical approach to the gravitational interaction, and it has the potential of opening a new window on the physical aspects of cosmology. In the present work we have presented some of the basic tools and theoretical concepts, which may be used for the further development of this theory.

\begin{acknowledgments}
We would like to thank the two anonymous reviewers for comments and suggestions that helped us to significantly improve our manuscript. The work of TH is supported by a grant of the Romanian Ministry of Education and Research, CNCS-UEFISCDI, project number PN-III-P4-ID-PCE-2020-2255 (PNCDI III). FSNL acknowledges support from the Funda\c{c}\~{a}o para a Ci\^{e}ncia e a Tecnologia (FCT) Scientific Employment Stimulus contract with reference CEECINST/00032/2018, and funding from the research grant CERN/FIS-PAR/0037/2019.
RACC, FSNL and MASP acknowledge support from the FCT research grants UIDB/04434/2020 and UIDP/04434/2020, and through the FCT project with reference PTDC/FIS-AST/0054/2021 (``BEYond LAmbda'').
MASP also acknowledges support from the FCT through the Fellowship UI/BD/154479/2022.
JLR was supported by the European Regional Development Fund and the programme Mobilitas Pluss (MOBJD647) and project No.~2021/43/P/ST2/02141 co-funded by the Polish National Science Centre and the European Union Framework Programme for Research and Innovation Horizon 2020 under the Marie Sklodowska-Curie grant agreement No. 94533.
\end{acknowledgments}


\end{document}